\documentclass[11pt]{article}

\usepackage{epsfig}
\usepackage{theorem}
\usepackage{enumerate}
\usepackage{amsmath}
\usepackage{amssymb}
\usepackage{boxedminipage}
\usepackage{subfigure}
\usepackage{url}
\usepackage{color}
\usepackage{epsfig}

\definecolor{myblue}{rgb}{0.165,0.34,0.5}

\newcommand{\titlestring}{Analysis of Evolutionary Algorithms on the One-Dimensional Spin Glass with Power-Law Interactions}

\newcommand{\reportnumber}{2009004}
\newcommand{\shortauthors}{Martin Pelikan and Helmut G. Katzgraber}
\newcommand{\datestring}{January 2009}
\date{}

\begin{document}


\begin{titlepage}
\setlength{\parindent}{0pt}

\noindent
\includegraphics[width=5in]{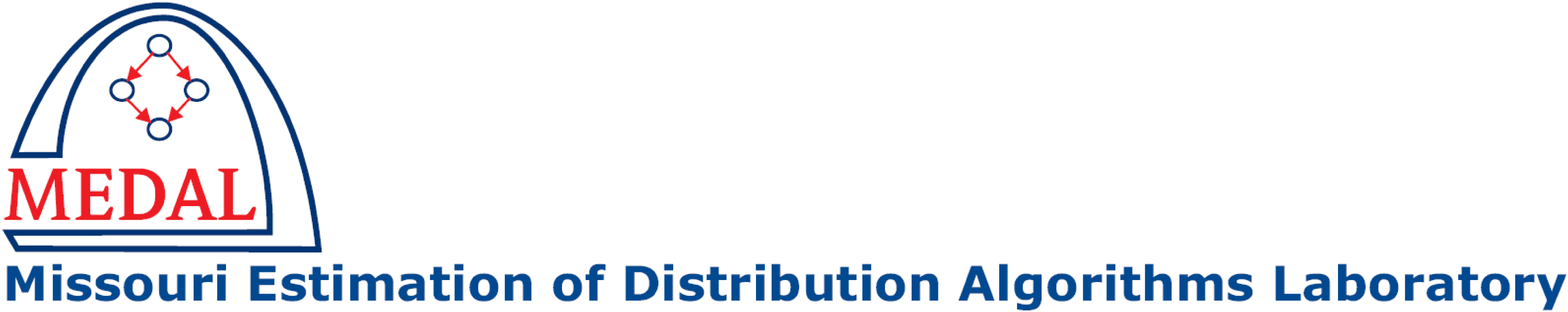}
\vspace*{0.075in}
{\color{myblue}
\hrule height 2pt
}
\vspace*{0.5in}

{\bf
\textsf{{\large
\titlestring}}
}

\vspace*{0.25in}

\textsf{\shortauthors}

\vspace*{0.25in}

\textsf{MEDAL Report No. \reportnumber}

\vspace*{0.25in}

\textsf{\datestring}

\vspace*{0.25in}

{\bf \textsf{Abstract}}  

\vspace*{0.075in}

{\small \textsf{This paper provides an in-depth empirical analysis of several evolutionary algorithms on the one-dimensional spin glass model with power-law interactions. The considered spin glass model provides a mechanism for tuning the effective range of interactions, what makes the problem interesting as an algorithm benchmark. As algorithms, the paper considers the genetic algorithm (GA) with twopoint and uniform crossover, and the hierarchical Bayesian optimization algorithm (hBOA). hBOA is shown to outperform both variants of GA, whereas GA with uniform crossover is shown to perform worst. The differences between the compared algorithms become more significant as the problem size grows and as the range of interactions decreases. Unlike for GA with uniform crossover, for hBOA and GA with twopoint crossover, instances with short-range interactions are shown to be easier. The paper also points out interesting avenues for future research.}}

\vspace*{0.25in}

{\bf \textsf{Keywords}}

\vspace*{0.075in}
{\small \textsf{Spin glass, power-law interactions, hierarchical BOA, genetic algorithm, estimation of distribution algorithms, evolutionary computation.}}

\vfill

\noindent
\begin{minipage}{6in}
{\small \textsf{Missouri Estimation of Distribution Algorithms Laboratory (MEDAL)\\
Department of Mathematics and Computer Science\\
University of Missouri--St. Louis\\
One University Blvd.,
St. Louis, MO 63121\\
E-mail: \url{medal@cs.umsl.edu}\\
WWW: \url{http://medal.cs.umsl.edu/}\\}}
\end{minipage}

\end{titlepage}


\begin{sloppy}

\title{\titlestring}

\author{
Martin Pelikan\\
Missouri Estimation of Distribution Algorithms Laboratory (MEDAL)\\
Dept. of Mathematics and Computer Science\\
Univ. of Missouri in St. Louis\\
One University Blvd., St. Louis, MO 63121\\
\url{pelikan@cs.umsl.edu}
\and
Helmut G. Katzgraber\\
Theoretische Physik,
ETH Zurich,
CH-8093 Zurich, Switzerland;\\
Dept. of Physics,
Texas A\&M University,
College Station, TX 77843-4242\\
\url{katzgraber@phys.ethz.ch}
}

\maketitle


\begin{abstract}
This paper provides an in-depth empirical analysis of several evolutionary algorithms on the one-dimensional spin glass model with power-law interactions. The considered spin glass model provides a mechanism for tuning the effective range of interactions, what makes the problem interesting as an algorithm benchmark. As algorithms, the paper considers the genetic algorithm (GA) with twopoint and uniform crossover, and the hierarchical Bayesian optimization algorithm (hBOA). hBOA is shown to outperform both variants of GA, whereas GA with uniform crossover is shown to perform worst. The differences between the compared algorithms become more significant as the problem size grows and as the range of interactions decreases. Unlike for GA with uniform crossover, for hBOA and GA with twopoint crossover, instances with short-range interactions are shown to be easier. The paper also points out interesting avenues for future research.
\end{abstract}



\section{Introduction}

Spin glasses are prototypical models for disordered systems, which provide a rich source of difficult computational problems. From the optimization perspective, the problem of finding ground states of various spin glass models represents a popular class of challenging optimization problems useful for analysis, design, and enhancement of optimization algorithms. There are a variety of spin glass models available, from short-range models arranged on a 2D lattice to the infinitely dimensional Sherrington-Kirkpatrick (SK) model~\cite{Sherrington:78}. One of the interesting models is the one-dimensional spin glass model with power-law interactions~\cite{kotliar:83,fisher:88} where the range of interactions can be controlled via a parameter. This provides the user with enough flexibility to cover a broad spectrum of spin glass models from short-range models, such as the one-dimensional nearest-neighbor spin glass, to long-range models, such as the infinitely dimensional SK spin glass. The one-dimensional model with power-law interactions has been recently used to elucidate some spin glass properties and suggested as an interesting algorithm benchmark~\cite{katzgraber:03,katzgraber:04c,Katzgraber:08}. Although spin glasses have been extensively used as test problems for evolutionary algorithms, most studies focused on short-range and finite dimensional systems such as the 2D or 3D spin glass~\cite{Muhlenbein:98a,Naudts:98*,Hartmann:01,Hoyweghen:01a,Pelikan:03*,Wegener:2004,Pelikan:06c}. Furthermore, the one-dimensional spin glass with power-law interactions has not yet been studied in the context of evolutionary algorithms.

This paper presents an in-depth empirical analysis of several evolutionary algorithms on the one-dimensional spin glass model with power-law interactions. In the comparison, the paper considers the genetic algorithm with two crossover operators and the hierarchical Bayesian optimization algorithm. The effects of the parameter that controls the range of interactions are studied and for each setting of this parameter a range of problem sizes are analyzed. The results are analyzed and related to known properties of the evolutionary algorithms under consideration. Important topics for future work in this area are also outlined. 

The paper is organized as follows. Section~\ref{section-1d-spin-glass} describes the one-dimensional spin glass model with power-law interactions. Section~\ref{section-algorithms} outlines evolutionary algorithms included in this comparison. Section~\ref{section-experiments} presents and discusses experimental results. Section~\ref{section-future-work} focuses on interesting topics for future work in this area. Finally, section~\ref{section-conclusions} summarizes and concludes the paper.


\section{One-Dimensional Spin Glass with Power-Law Interactions}
\label{section-1d-spin-glass}
The Sherrington-Kirkpatrick spin glass~\cite{Sherrington:78} is described by a set of spins $\{s_i\}$ and a set of couplings $\{J_{i,j}\}$ between all pairs of spins. 
For the classical Ising
model, each spin $s_i$ can be in one of two states: $s_i=+1$ or 
$s_i=-1$. This simplification corresponds to highly 
anisotropic systems; nevertheless, the two-state 
Ising model comprises all basic effects also
found in models with more degrees of freedom.

For a set of coupling constants $\{J_{i,j}\}$, and a configuration of spins $C=\{s_i\}$, the energy can be
computed as

\begin{equation}
H(C) = - \sum_{i<j} J_{i,j} s_i s_j.
\end{equation}

There are two typical tasks in statistical physics: (1) integrate a known function over all possible configurations of spins for given coupling constants assuming the Boltzmann distribution of spin configurations; (2) find one ore more ground states (spin configurations associated with the minimum possible energy). The problem of finding ground states is NP-complete even when the interactions are limited only to neighbors in a 3D lattice~\cite{Barahona:82}; the SK spin glass is thus certainly NP-complete unless we severely restrict couplings. 

Typically, one would study a number of randomly generated spin glass instances where couplings for each instance would be generated according to the Gaussian distribution with zero mean and unit variance. In the one-dimensional spin glass model with power-law interactions~\cite{Katzgraber:08}, spins are arranged equidistantly on a circle with circumference $n$. The spins are numbered in a counterclockwise fashion. 
Every spin interacts with every other spin, just like in the standard SK spin glass; however, interactions between spins located further from each other are weaker, and the effects of distance can be controlled with a parameter, providing a mechanism for tuning the effective range of interactions. More specifically, the couplings are generated according to 

\begin{equation}
J_{i,j} = c(\sigma) \frac{\epsilon_{i,j}}{r_{i,j}^\sigma},
\end{equation}

\noindent
where $\epsilon_{i,j}$ are generated according to normal distribution with zero mean and unit variance, $c(\sigma)$ is a normalization constant, $\sigma>0$ is the user-specified parameter to control the effective range of interactions, and $r_{i,j}=n \sin (\pi |i-j|/n) / \pi$ denotes the geometric distance between $s_i$ and $s_j$ (see figure~\ref{fig-sk-ring}). The magnitude of spin-spin couplings decreases with their distance. Furthermore, as discussed shortly, the effects of distance on the magnitude of couplings increase with $\sigma$.

By varying $\sigma$ one can tune the model from the infinite-range
to the short-range universality class: For $0 < \sigma \le 1/2$ the
model is in the infinite-range universality in the sense that $\sum_j
[J_{ij}^2]_{\rm av}$ diverges, and for $\sigma = 0$ it corresponds
to the SK model.  For $1/2 < \sigma
\le 2/3$ the model describes a mean-field long-range spin glass,
corresponding---in the analogy with short-range systems---to a
short-range model in dimension above the upper critical dimension
$d \ge d_{\rm u} = 6$. For $2/3 < \sigma \le 1$ the model has
non-mean-field critical behavior with a finite transition temperature
$T_c$, and for $\sigma \ge 1$, the transition temperature is zero
and the behavior of the model is short ranged.

\begin{figure}
\centering
\epsfig{file=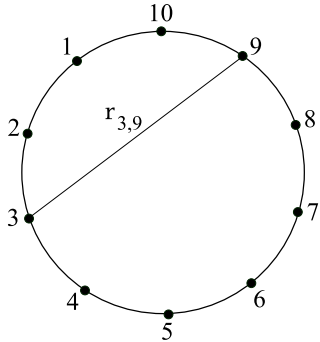,width=0.25 \textwidth}
\caption{One-dimensional spin glass of size $n=10$ arranged on a ring.}
\label{fig-sk-ring}
\end{figure}

To find guaranteed ground states, a branch-and-bound algorithm adopted from refs.~\cite{Kobe:84,Kobe:03} was used. This allows us to find guaranteed ground states of instances of up to about $n=100$ spins for larger values of $\sigma$ and up to about $n=80$ spins for smaller values of $\sigma$. For larger systems, we used the population-doubling approach proposed in ref.~\cite{Pelikan:08}.


\section{Compared Algorithms}
\label{section-algorithms}
The genetic algorithm (GA)~\cite{Holland:75a,Goldberg:89d} evolves a population of candidate solutions typically represented by binary strings of fixed length with the first population generated at random according to the uniform distribution over all binary strings. Each iteration starts by selecting promising solutions from the current population; we use binary tournament selection without replacement. New solutions are created by applying variation operators to the population of selected solutions. Specifically, crossover is used to exchange bits and pieces between pairs of candidate solutions and mutation is used to perturb the resulting solutions. Here we use uniform or twopoint crossover, and bit-flip mutation~\cite{Goldberg:89d}. To maintain useful diversity in the population, the new candidate solutions are incorporated into the original population using restricted tournament selection (RTS)~\cite{Harik:95a}. The run is terminated when termination criteria are met. In this paper, each run is terminated either when the global optimum has been found or when a maximum number of iterations has been reached.

The hierarchical Bayesian optimization algorithm (hBOA)~\cite{Pelikan:01*,Pelikan:03b,Pelikan:book} is an estimation of distribution algorithm (EDA)~\cite{Baluja:94,Muhlenbein:96**,Larranaga:02,Pelikan:02,Larranaga:06,Pelikan:EDA-book}. EDAs---also called probabilistic model-building genetic algorithms (PMBGAs)~\cite{Pelikan:02} and iterated density estimation algorithms (IDEAs)~\cite{Bosman:00*}---differ from GAs by replacing standard variation operators of GAs such as crossover and mutation by building a probabilistic model of promising solutions and sampling the built model to generate new candidate solutions. The only difference between GA and hBOA variants used in this study is that instead of using crossover and mutation to create new candidate solutions, hBOA learns a Bayesian network with local structures~\cite{Chickering:97,Friedman:99} as a model of the selected solutions and generates new candidate solutions from the distribution encoded by this model. For more details on hBOA, see refs.~\cite{Pelikan:01*,Pelikan:book}.

The deterministic hill climber (DHC) is incorporated into both GA and hBOA to improve their performance similarly as in previous studies on using GA and hBOA for solving spin glasses~\cite{Pelikan:03*,Pelikan:book}. DHC takes a candidate solution represented by an $n$-bit binary string on input. Then, it performs one-bit changes on the solution that lead to the maximum improvement of solution quality. DHC is terminated when no single-bit flip improves solution quality and the solution is thus locally optimal. Here, DHC is used to improve every solution in the population before the evaluation is performed. 

To represent a configurations for $n$ spins, both hBOA and GA use $n$-bit binary strings with each bit corresponding to one of the spins; state $+1$ is represented by a $1$, state $-1$ is represented by a $0$. 


\section{Experiments}
\label{section-experiments}

\subsection{Problem Instances}
Instances of sizes $n=20$ to $n=150$ were considered, and for each problem size, the coefficient $\sigma$ was set to each of the values from $\{0.00,0.55,0.75,0.83,1.00,1.50,2.00\}$ (the set of values of $\sigma$ was adopted from ref.~\cite{Boettcher:08}). For each combination of $n$ and $\sigma$, 10,000 problem instances were generated and tested. This resulted in the total of 610,000 unique problem instances. Instances of up to $n=100$ were solved using the branch and bound algorithm; other instances were solved with the population-doubling method and hBOA as described in ref.~\cite{Pelikan:08}.

\subsection{Experimental Methodology}
For each problem instance and each algorithm, an adequate population size was approximated using bisection~\cite{Sastry:01c,Pelikan:book} to ensure that the optimum is found in 10 out of 10 independent runs. Each run was terminated when the global optimum was found (success) or when the maximum number of generations $n$ was reached (failure). The most important statistic relating to the overall complexity of each algorithm is the number of DHC flips, since this statistic combines all important statistics and can be consistently compared regardless of the used algorithm. Due to the variety of hardware used to perform the simulations, it was not possible to compare the actual running times.

\subsection{Results}
The growth of the number of evaluations and the number of DHC flips until optimum for hBOA, GA with twopoint crossover, and GA with uniform crossover is shown in figures~\ref{fig-evals} and~\ref{fig-flips}. As expected, both the number of evaluations as well as the number of flips grow with problem size, and a similar observation can be made for the adequate population size (see figure~\ref{fig-popsizes}) and the number of generations (results omitted). This result is not surprising because the more decision variables in the problem, the more difficult the problem becomes. What is somewhat more interesting is how the rate of growth of these statistics changes with $\sigma$, which controls the range of interactions, and how the effects of $\sigma$ change depending on the algorithm under consideration; this is the topic discussed in the following few paragraphs.

\begin{figure*}[t]
\hfill ~
\subfigure[hBOA]{\epsfig{file=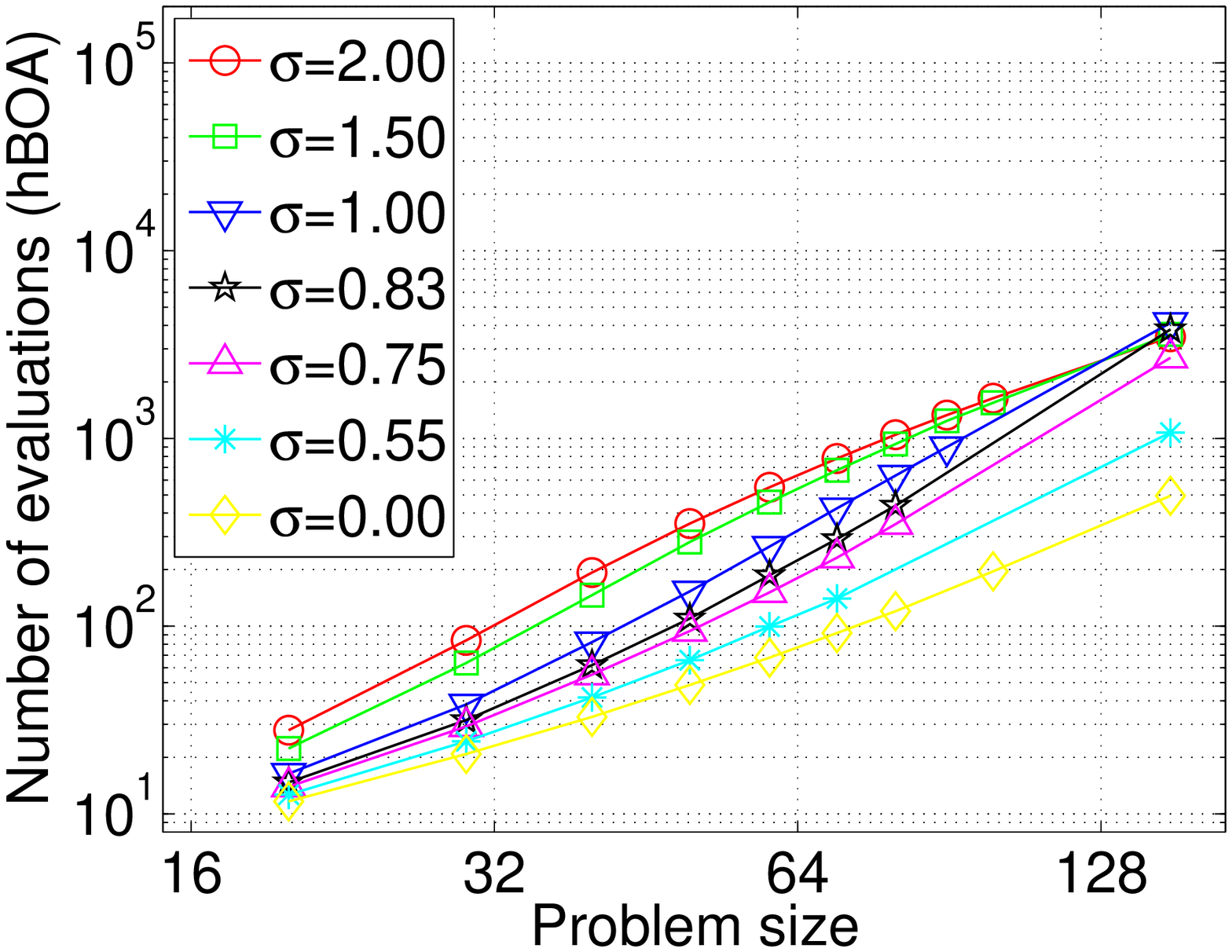,width=0.31 \textwidth}}
\hfill ~
\subfigure[GA (twopoint)]{\epsfig{file=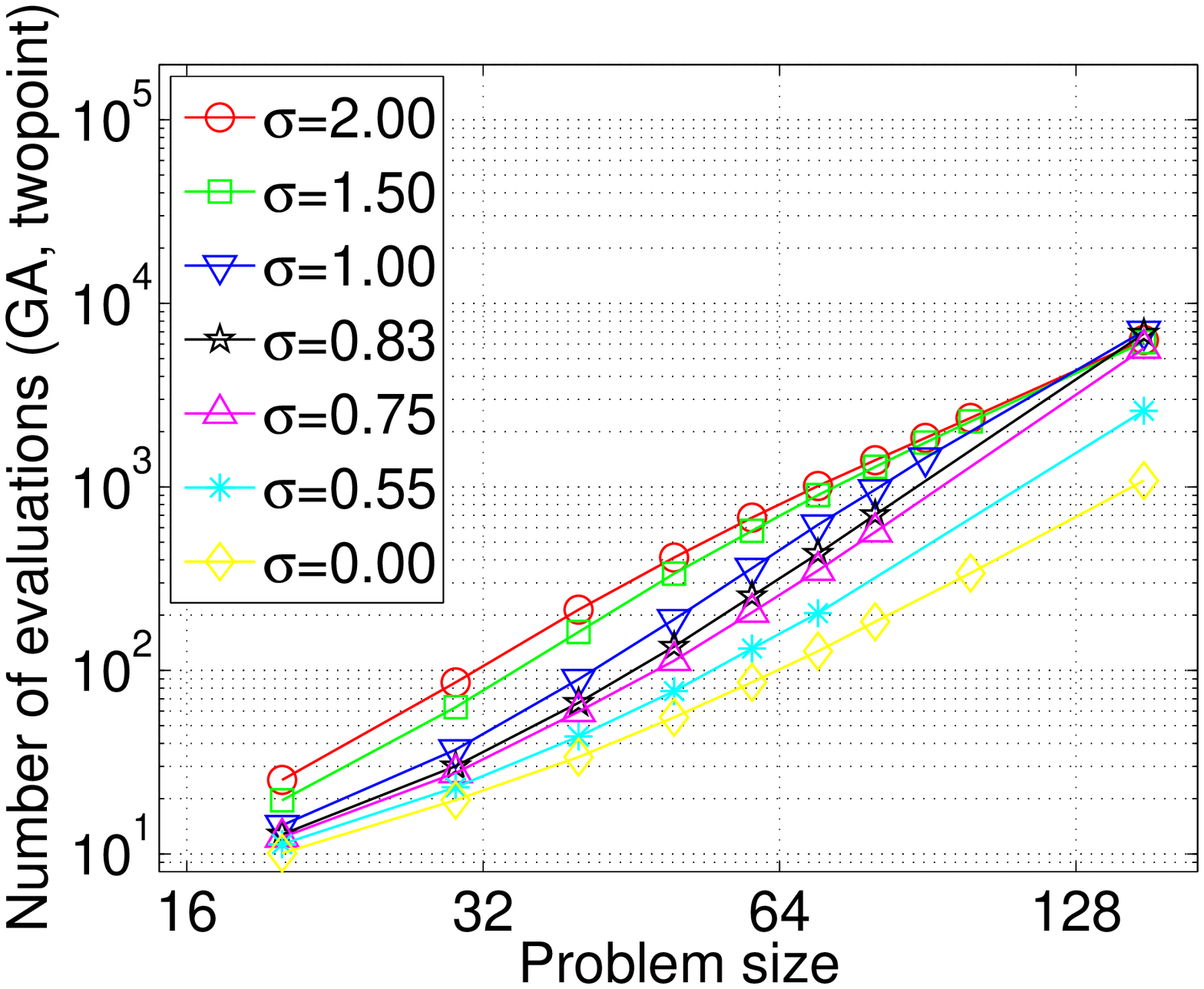,width=0.31 \textwidth}}
\hfill ~
\subfigure[GA (uniform)]{\epsfig{file=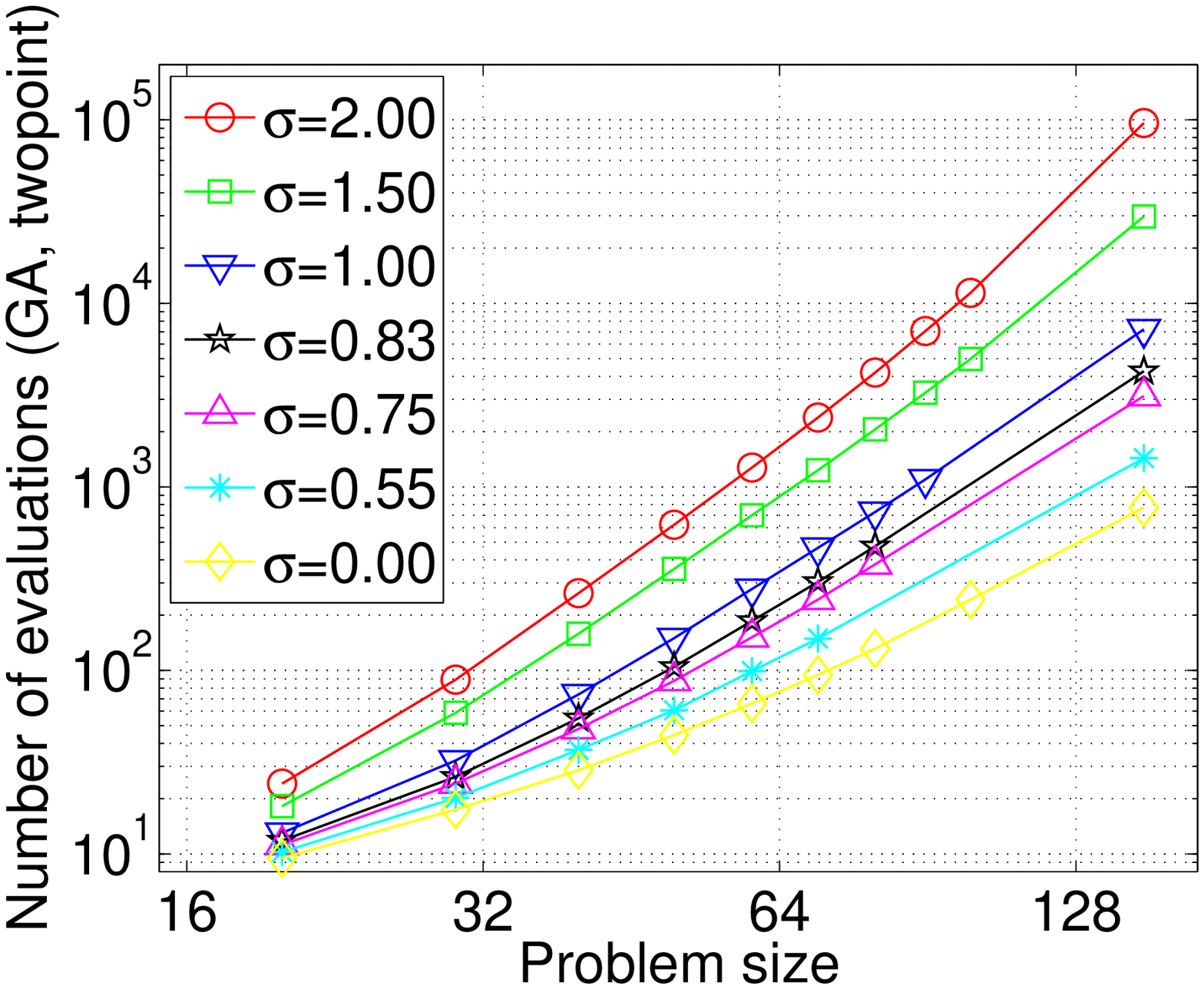,width=0.31 \textwidth}}
\hfill ~
\caption{Growth of the number of evaluations with problem size. }
\label{fig-evals}
\end{figure*}

\begin{figure*}[t]
\hfill ~
\subfigure[hBOA]{\epsfig{file=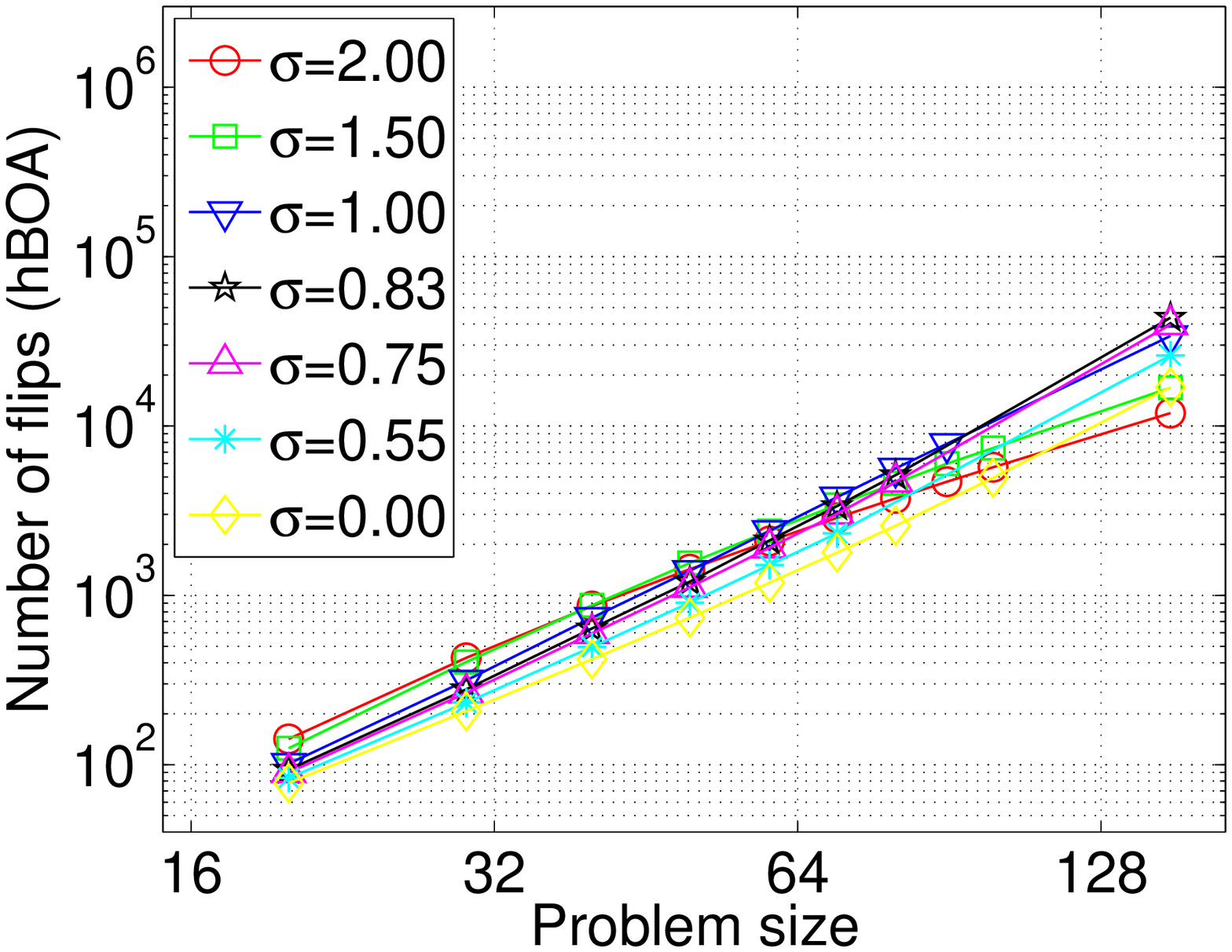,width=0.31 \textwidth}}
\hfill ~
\subfigure[GA (twopoint)]{\epsfig{file=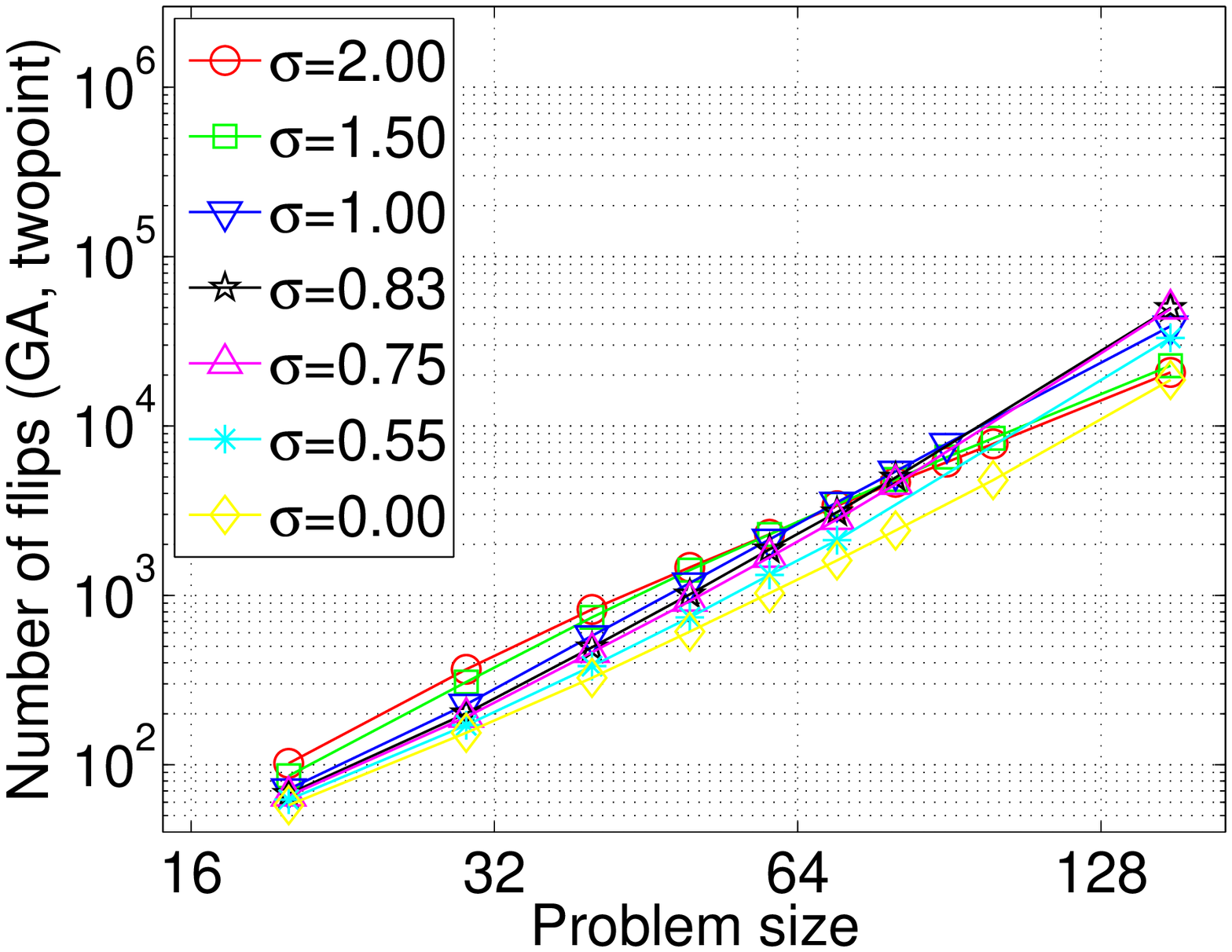,width=0.31 \textwidth}}
\hfill ~
\subfigure[GA (uniform)]{\epsfig{file=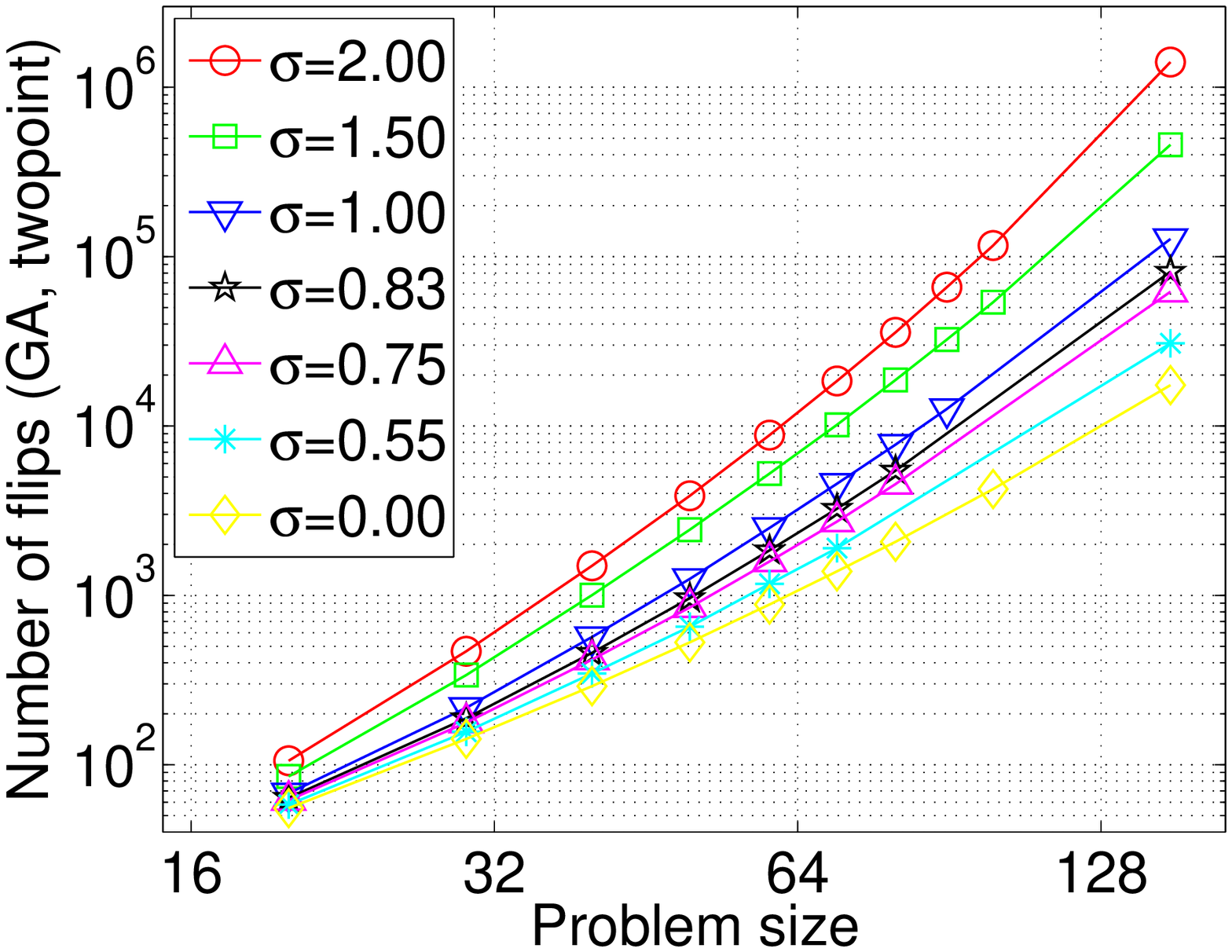,width=0.31 \textwidth}}
\hfill ~
\caption{Growth of the number of flips with problem size. }
\label{fig-flips}
\end{figure*}

\begin{figure*}[t]
\hfill ~
\subfigure[hBOA]{\epsfig{file=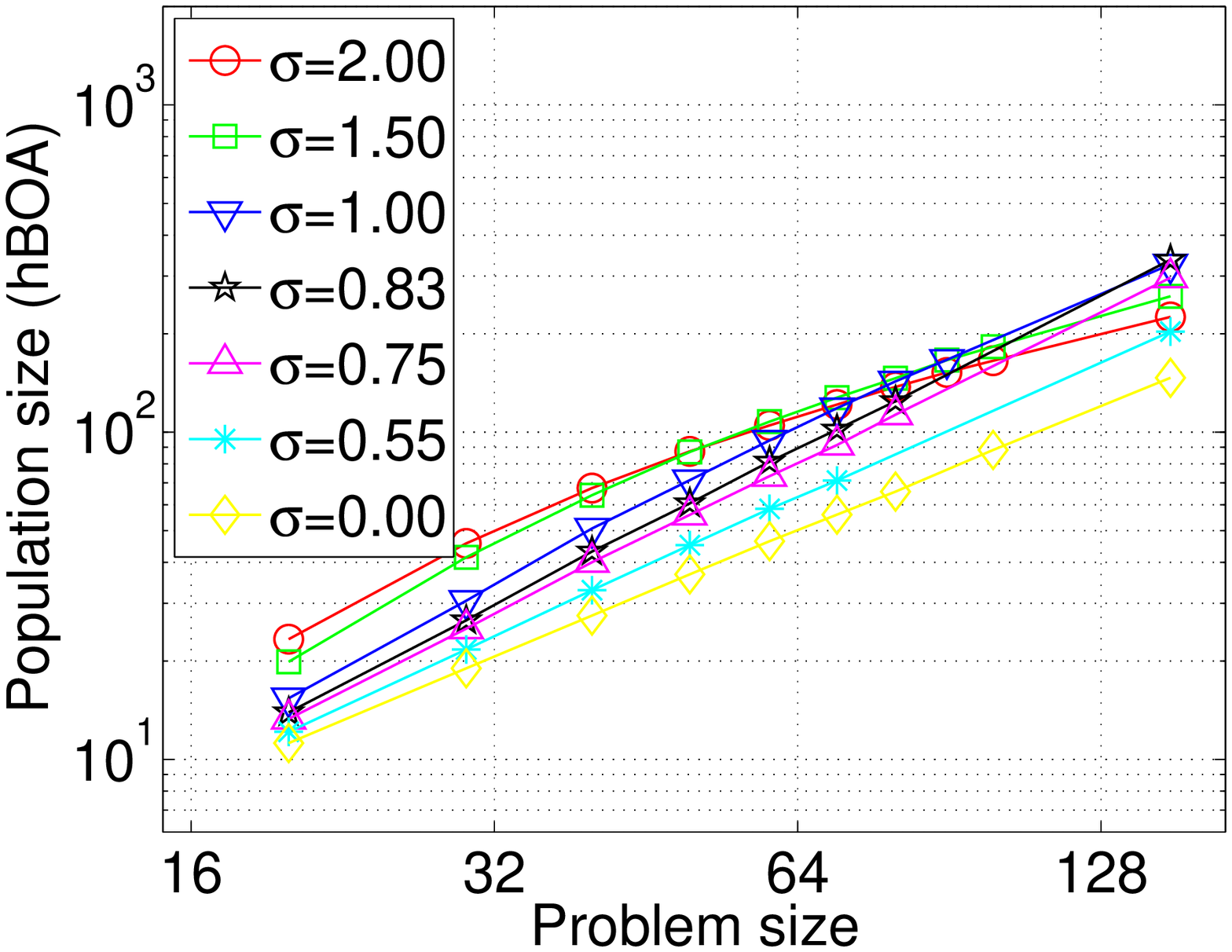,width=0.31 \textwidth}}
\hfill ~
\subfigure[GA (twopoint)]{\epsfig{file=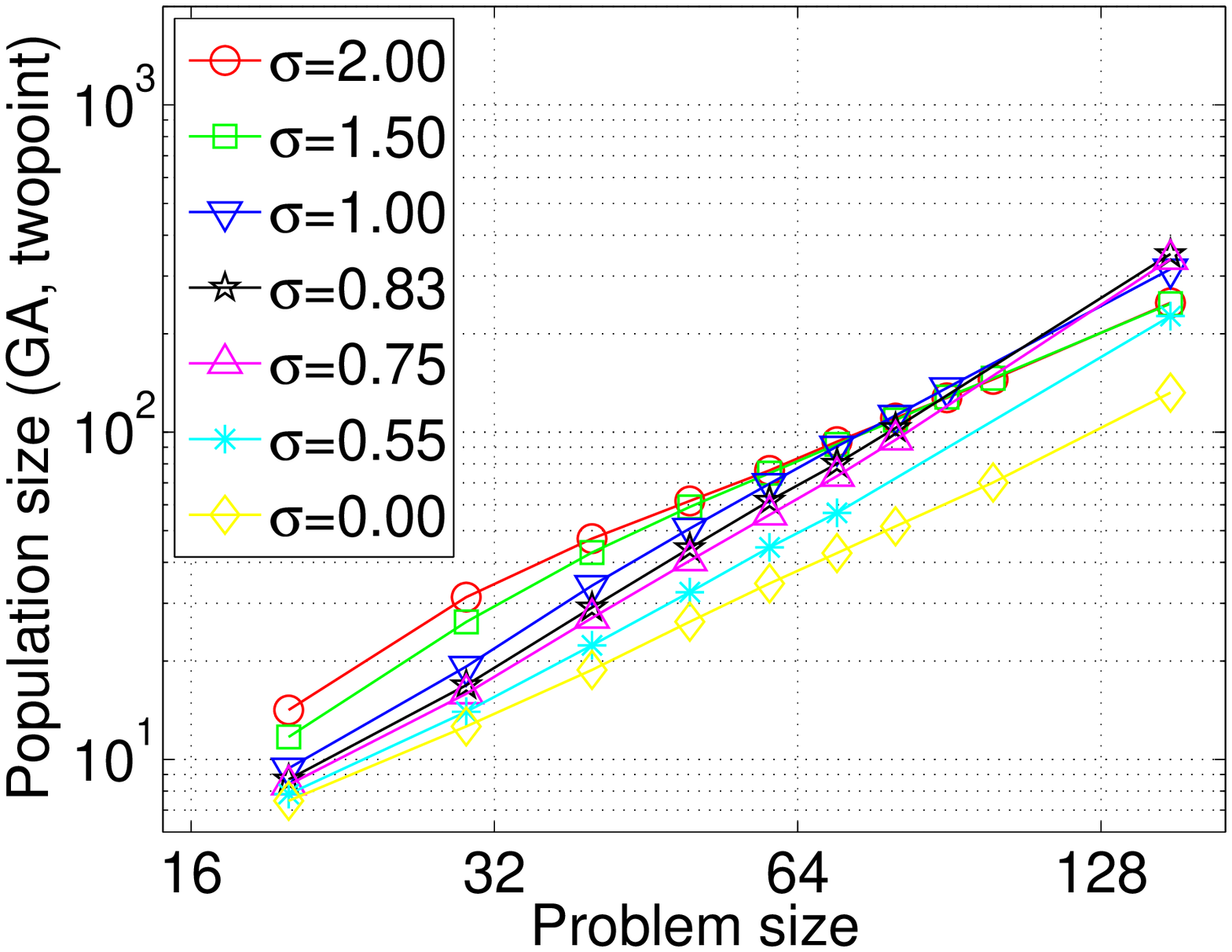,width=0.31 \textwidth}}
\hfill ~
\subfigure[GA (uniform)]{\epsfig{file=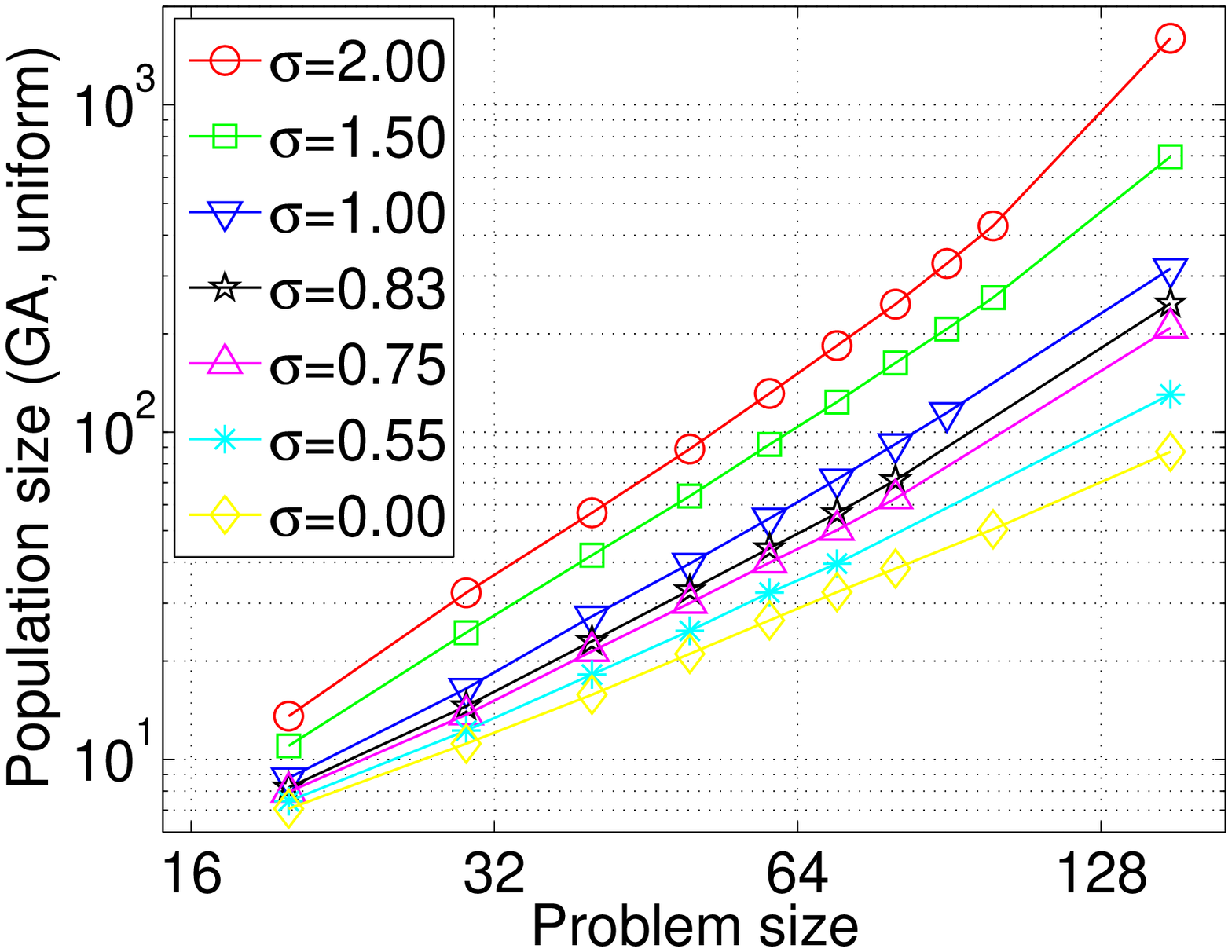,width=0.31 \textwidth}}
\hfill ~
\caption{Growth of the population size with problem size. }
\label{fig-popsizes}
\end{figure*}

Based on the definition of the 1D spin glass with power-law interactions, as the value of $\sigma$ grows, the range of the most significant interactions is reduced. With reduction of the range of interactions, the problem should become easier both for selectorecombinative GAs capable of linkage learning, such as hBOA, as well as for selectorecombinative GAs which rarely break interactions between closely located bits, such as GA with twopoint crossover. This is clearly demonstrated by the results for these two algorithms presented in figures~\ref{fig-evals} and~\ref{fig-flips}.  Although for many problem sizes, the absolute number of evaluations and the number of flips are in fact smaller for larger values of $\sigma$, the growth of the two statistics with problem size slows down substantially as $\sigma$ is increased. In fact, for the smallest values of $\sigma$, the number of evaluations and the number of flips both appear to grow faster than polynomially, whereas for the largest values of $\sigma$, the growth appears to be upper bounded by a polynomial of low order. On the other hand, for the uniform crossover, which does not respect interactions between consequent bits, the growth appears to be faster than polynomial for both the number of evaluations and the number of flips.

To better visualize the effects of $\sigma$ on the performance of compared algorithms, figures~\ref{fig-ratio200-evals} and~\ref{fig-ratio200-flips} show the ratio of the number of evaluations and the number of flips for $\sigma<2$ and $\sigma=2$. The larger the ratio, the better the results for $\sigma=2$ compared to smaller values of $\sigma$. For both hBOA and GA with twopoint crossover, the ratios for $\sigma\leq 1$ appear to grow relatively fast; in fact, the growth appears to be faster than polynomial; on the other hand, for $\sigma=1.5$, the growth can still be observed but it is very slow in both cases. For GA with uniform crossover, the opposite behavior can be observed; this indicates that uniform crossover does not benefit from the decrease in the range of interactions, at least within the range of values considered in this work. 

\begin{figure*}[t]
\hfill ~
\subfigure[hBOA]{\epsfig{file=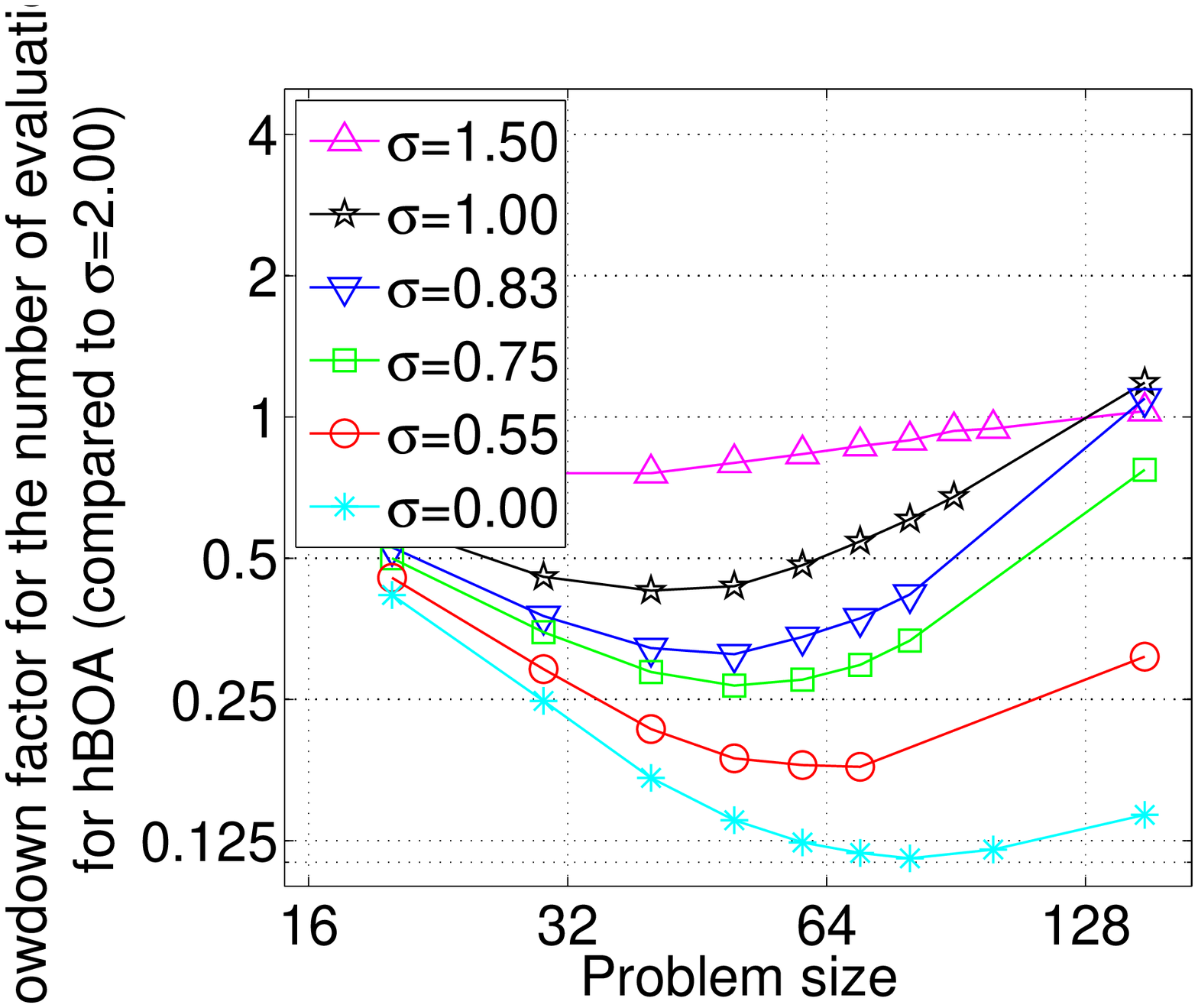,width=0.31 \textwidth}}
\hfill ~
\subfigure[GA (twopoint)]{\epsfig{file=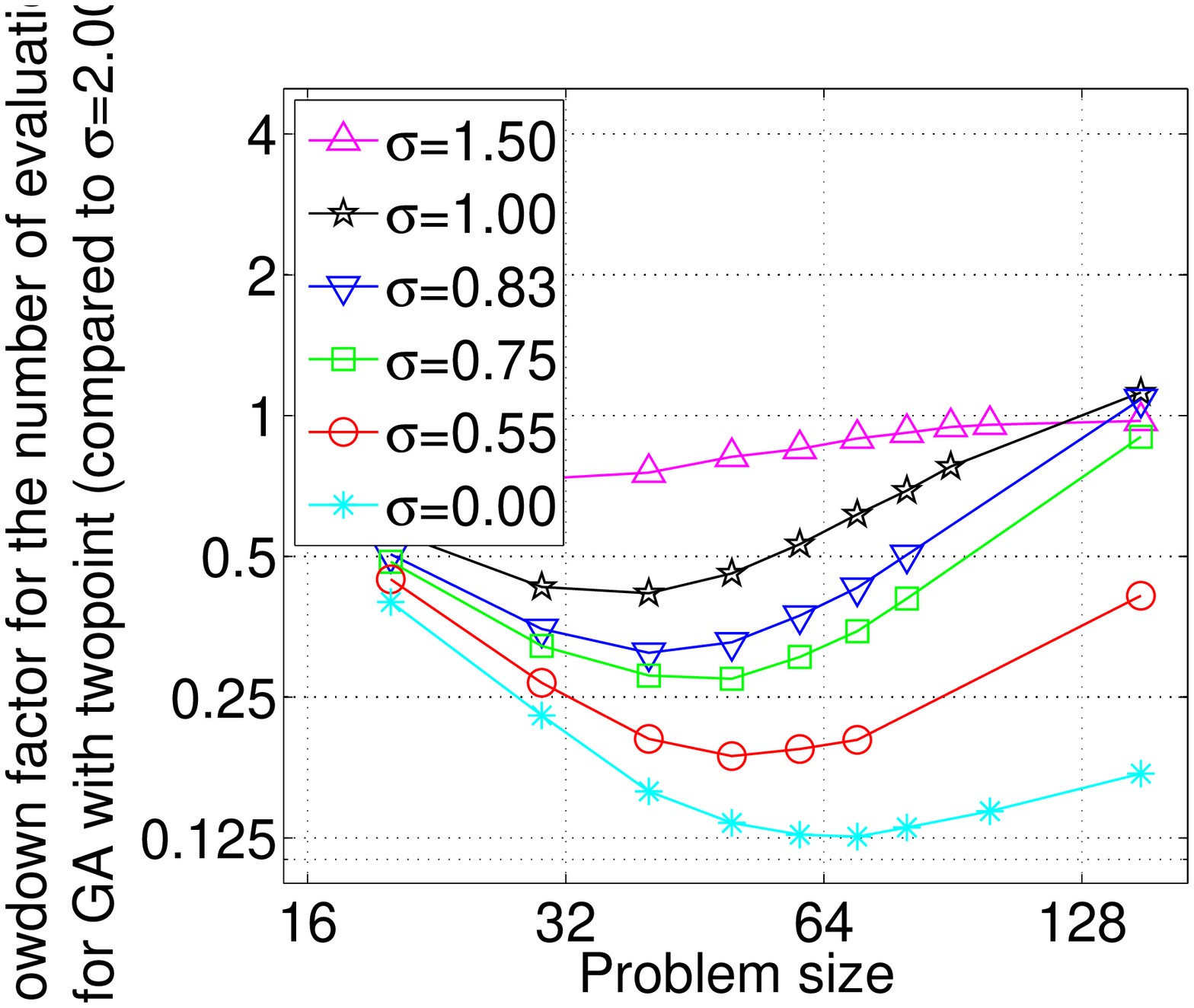,width=0.31 \textwidth}}
\hfill ~
\subfigure[GA (uniform)]{\epsfig{file=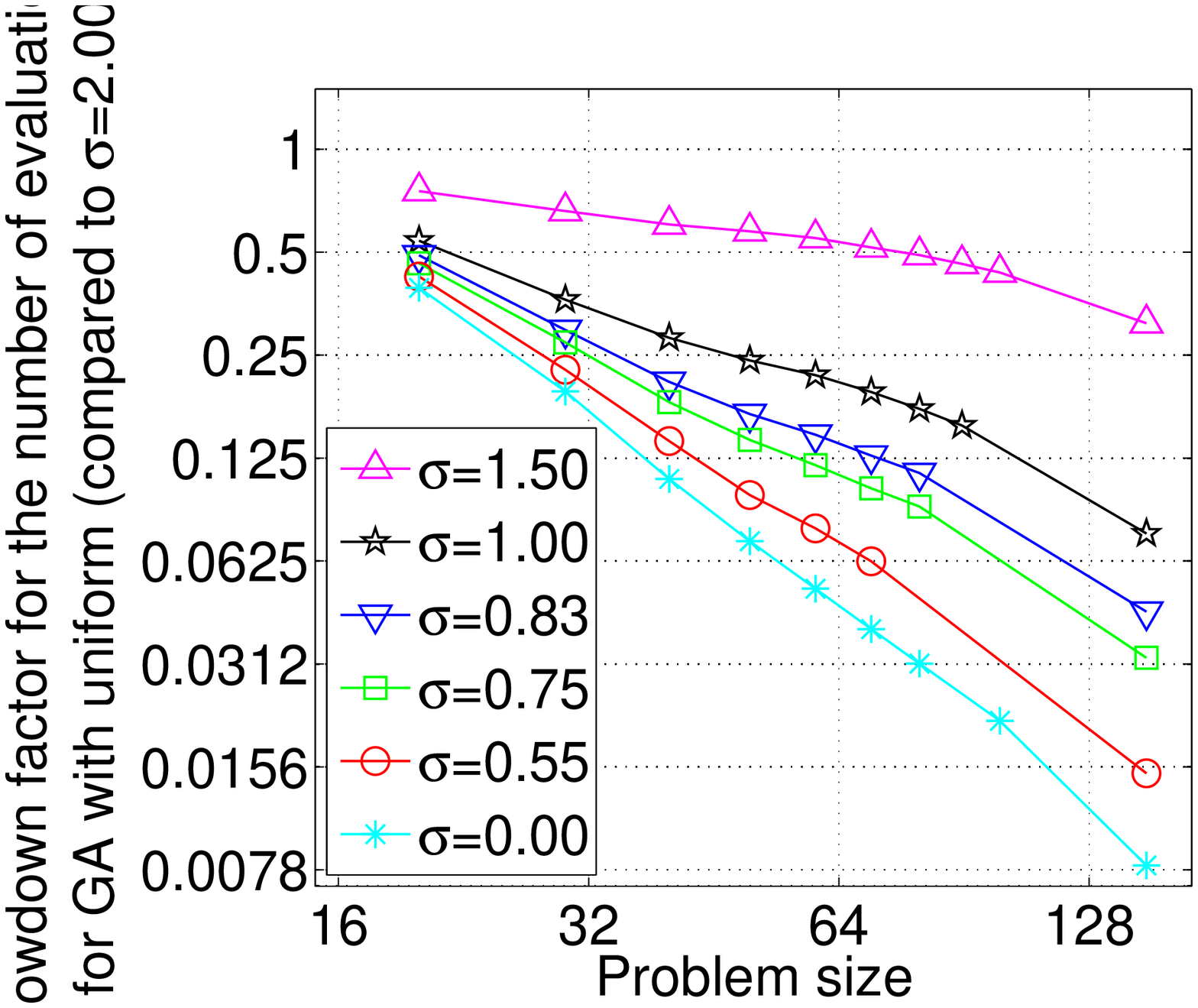,width=0.31 \textwidth}}
\hfill ~
\caption{Comparison of the number of evaluations for $\sigma=2.00$ with that for $\sigma<2.00$.}
\label{fig-ratio200-evals}
\end{figure*}

\begin{figure*}[t]
\hfill ~
\subfigure[hBOA]{\epsfig{file=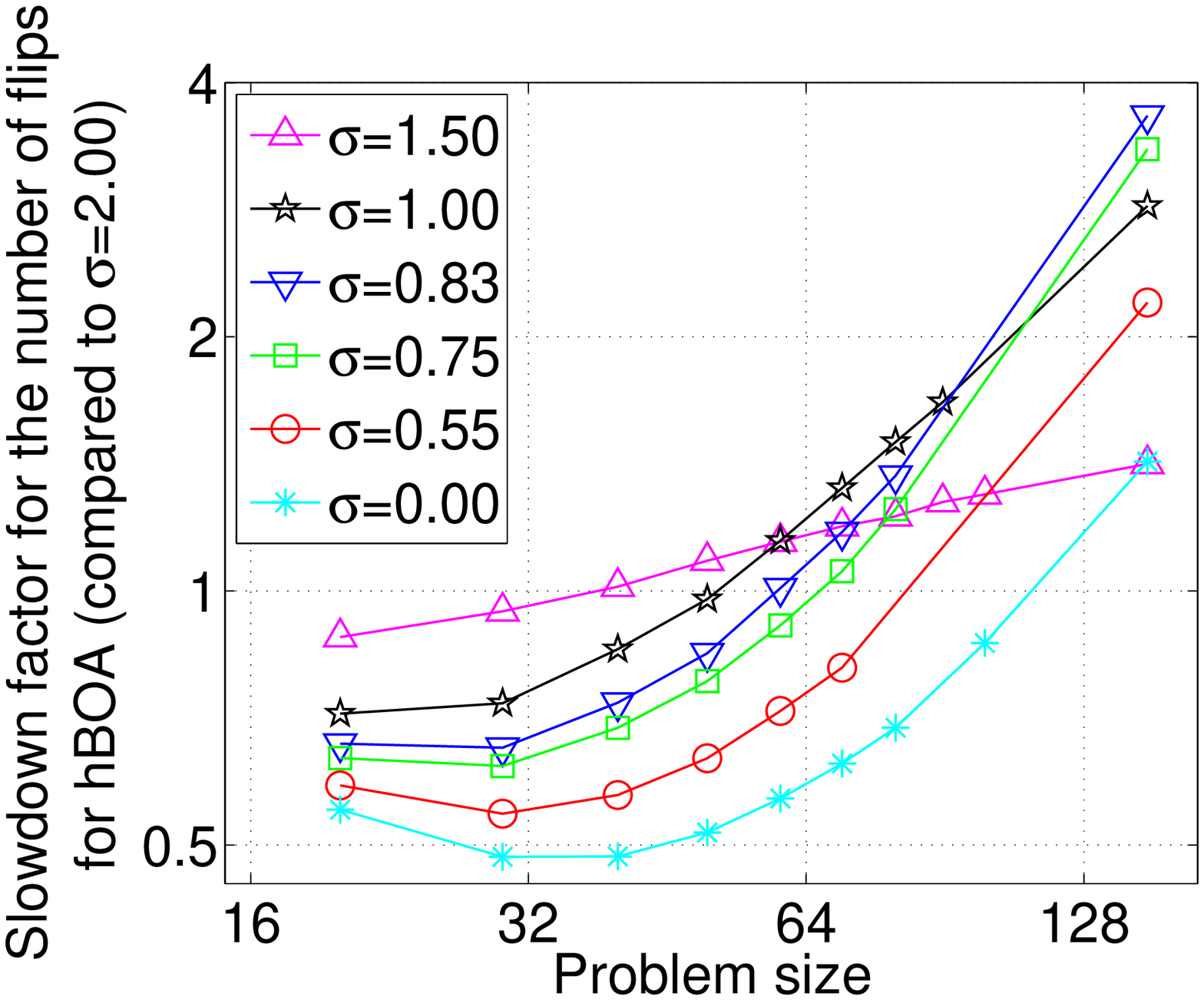,width=0.31 \textwidth}}
\hfill ~
\subfigure[GA (twopoint)]{\epsfig{file=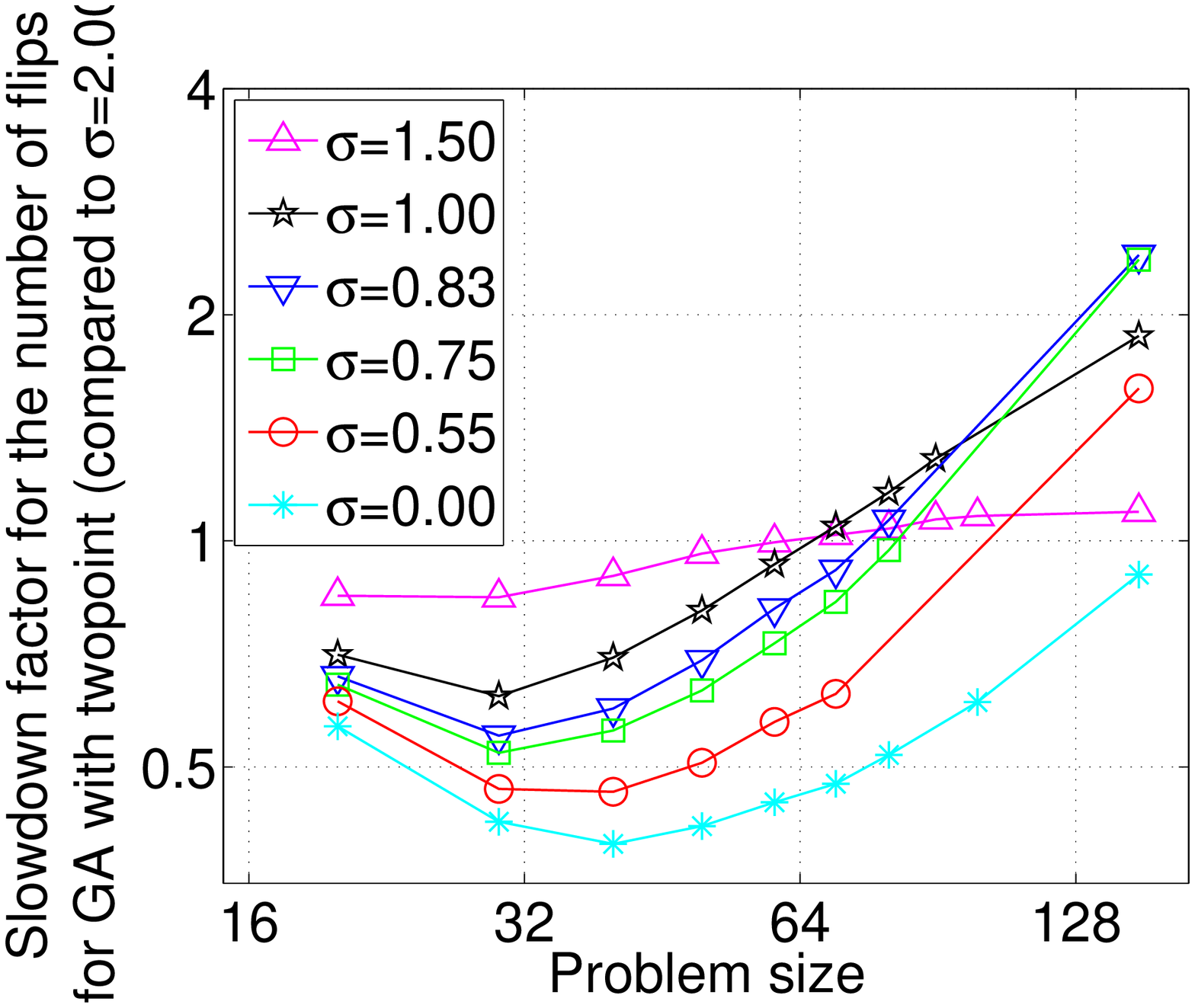,width=0.31 \textwidth}}
\hfill ~
\subfigure[GA (uniform)]{\epsfig{file=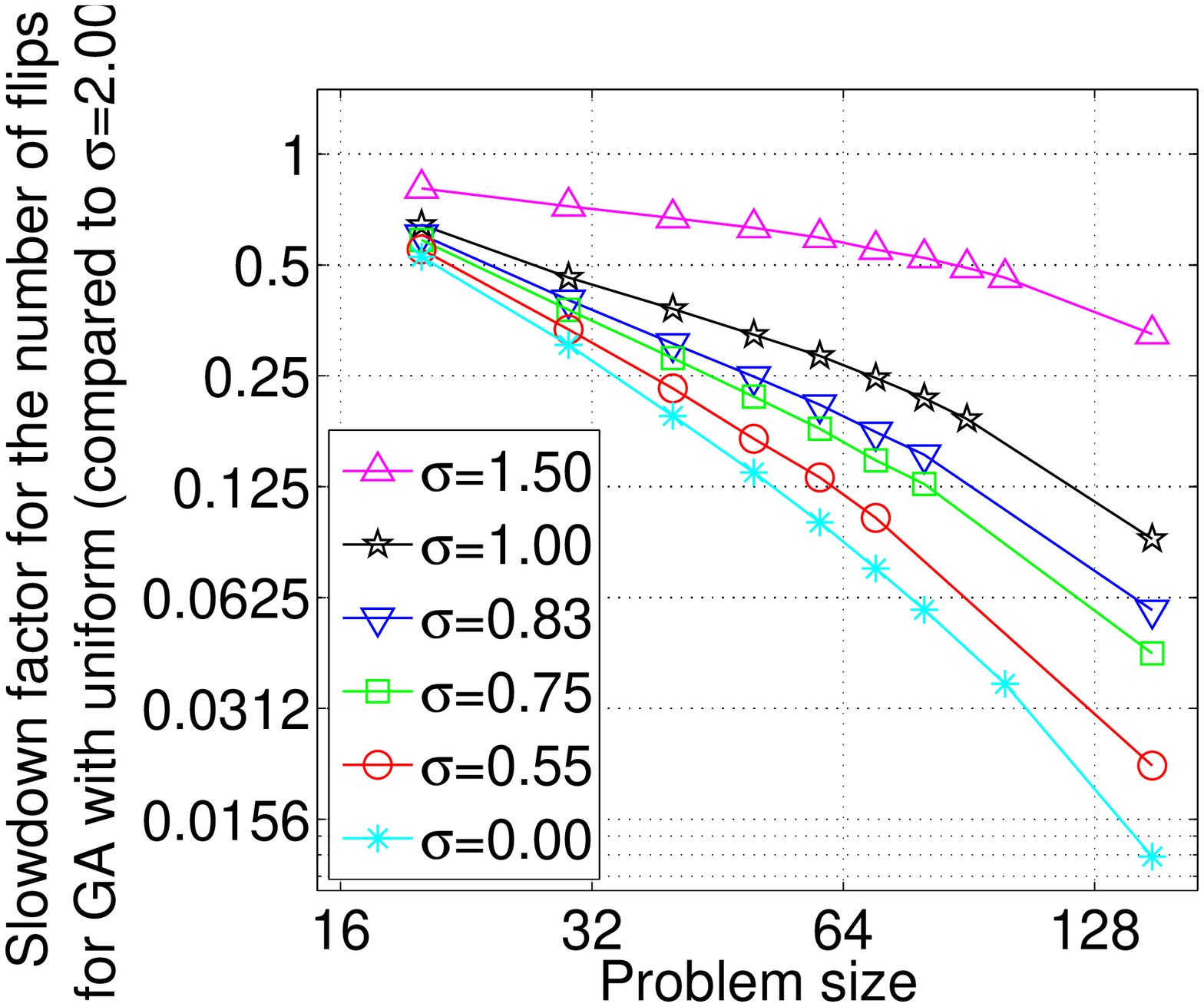,width=0.31 \textwidth}}
\hfill ~
\caption{Comparison of the number of flips for $\sigma=2.00$ with that for $\sigma<2.00$.}
\label{fig-ratio200-flips}
\end{figure*}

To compare performance of two algorithms on any specific problem instance, one may compute the ratio of the number of evaluations required by the two algorithms; analogical comparison can be performed for the number of flips, the population size, or the number of generations. These ratios can then be averaged over all instances for any considered combination of values of $n$ and $\sigma$. The comparison for the number of evaluations via such ratios is shown in figure~\ref{fig-ratio-evaluations} and analogical results for the number of flips and the required population sizes are shown in figures~\ref{fig-ratio-flips} and~\ref{fig-ratio-popsizes}, respectively. Pairwise comparisons of hBOA and GA show that with respect to both the number of evaluations as well as the number of flips, the best performance is achieved by hBOA, the second best performance is achieved by GA with twopoint crossover, and the worst performance is achieved by GA with uniform crossover. The differences between all algorithms increase with problem size and the differences for larger values of $\sigma$ are more significant than those for smaller values of $\sigma$; in other words, the shorter the range of interactions, the more significant the differences.

\begin{figure*}[t]
\hfill ~
\subfigure[GA (twopoint) and hBOA]{\epsfig{file=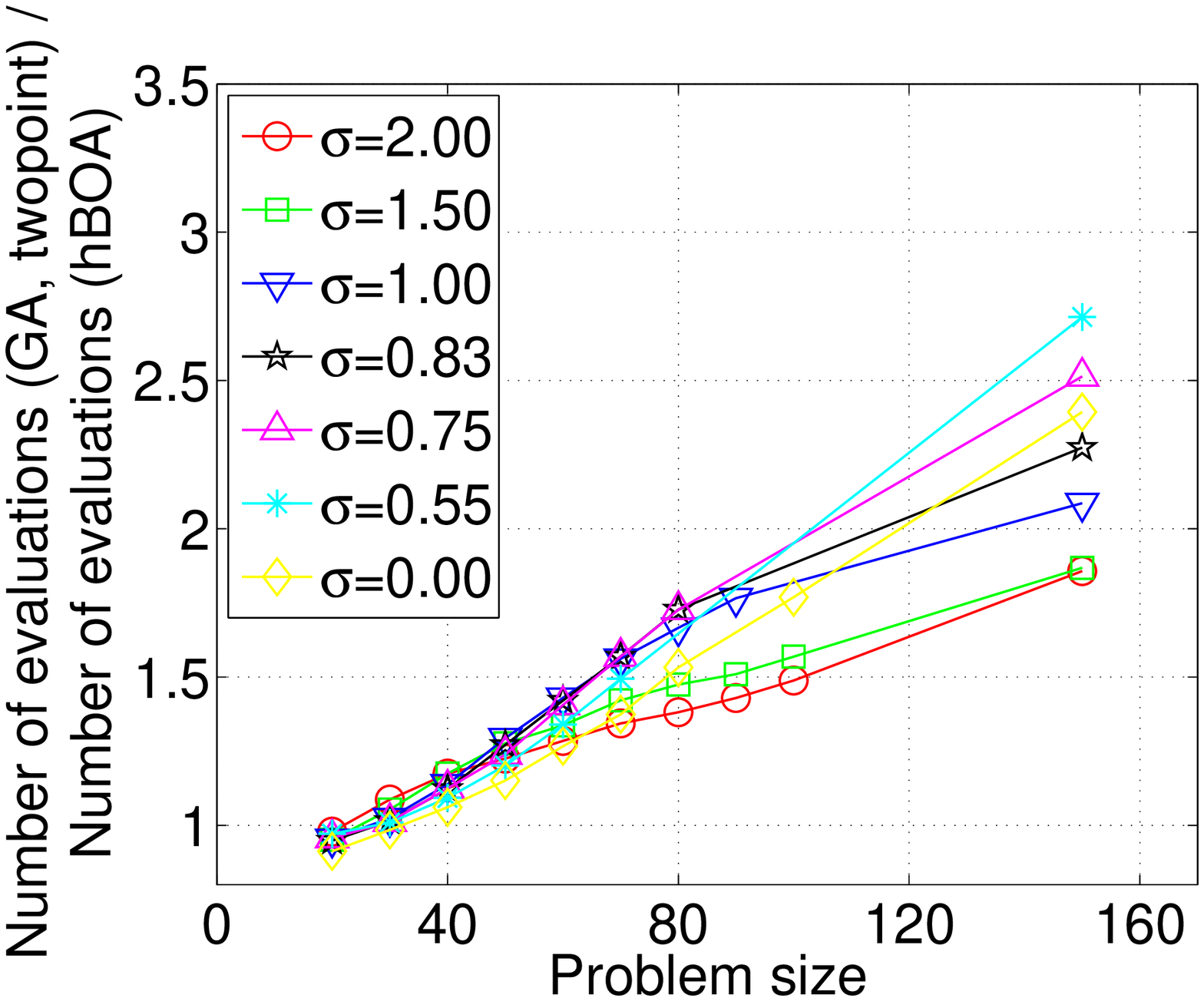,width=0.31 \textwidth}}
\hfill ~
\subfigure[GA (uniform) and hBOA]{\epsfig{file=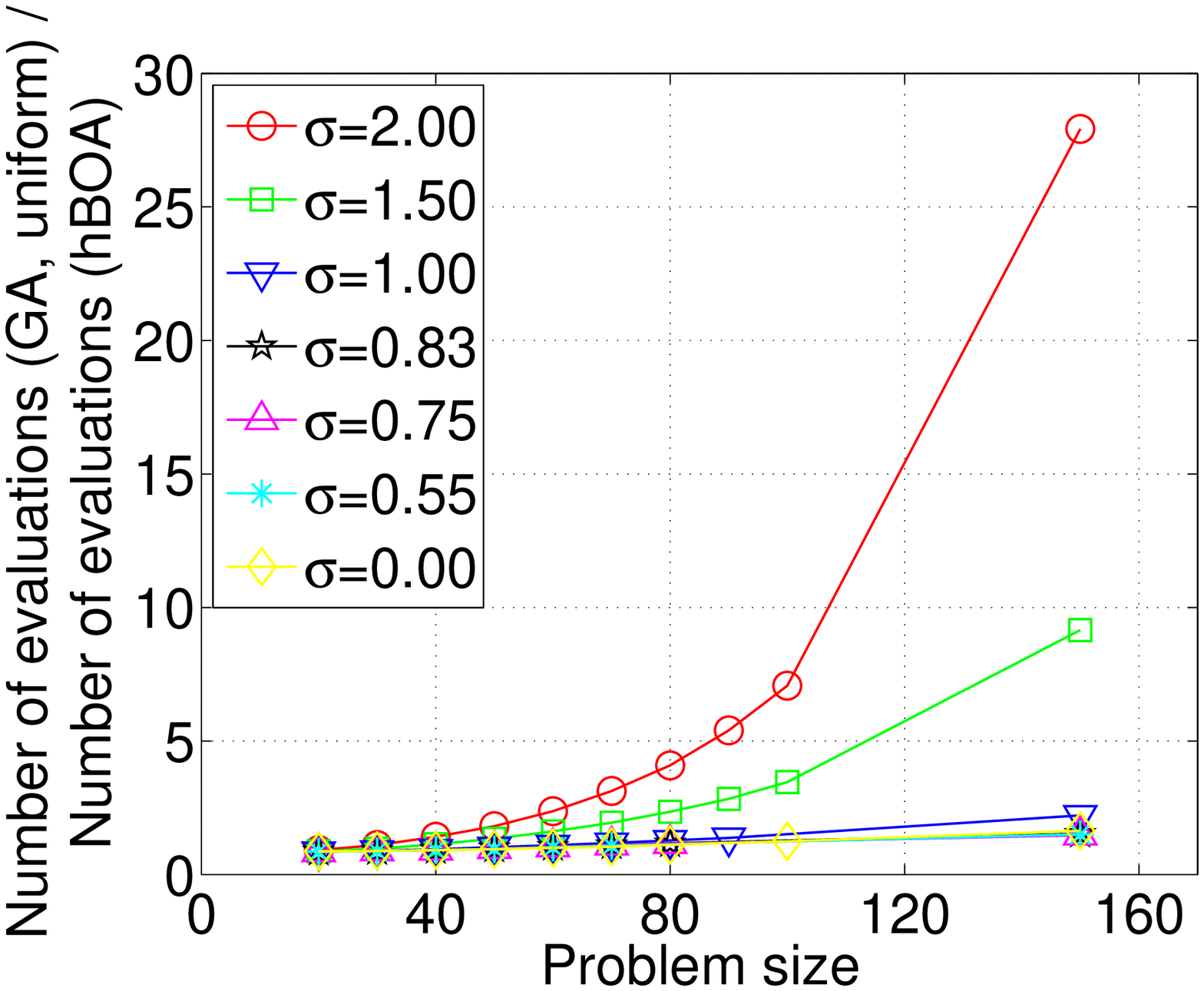,width=0.31 \textwidth}}
\hfill ~
\subfigure[GA (uniform) and GA (twopoint)]{\epsfig{file=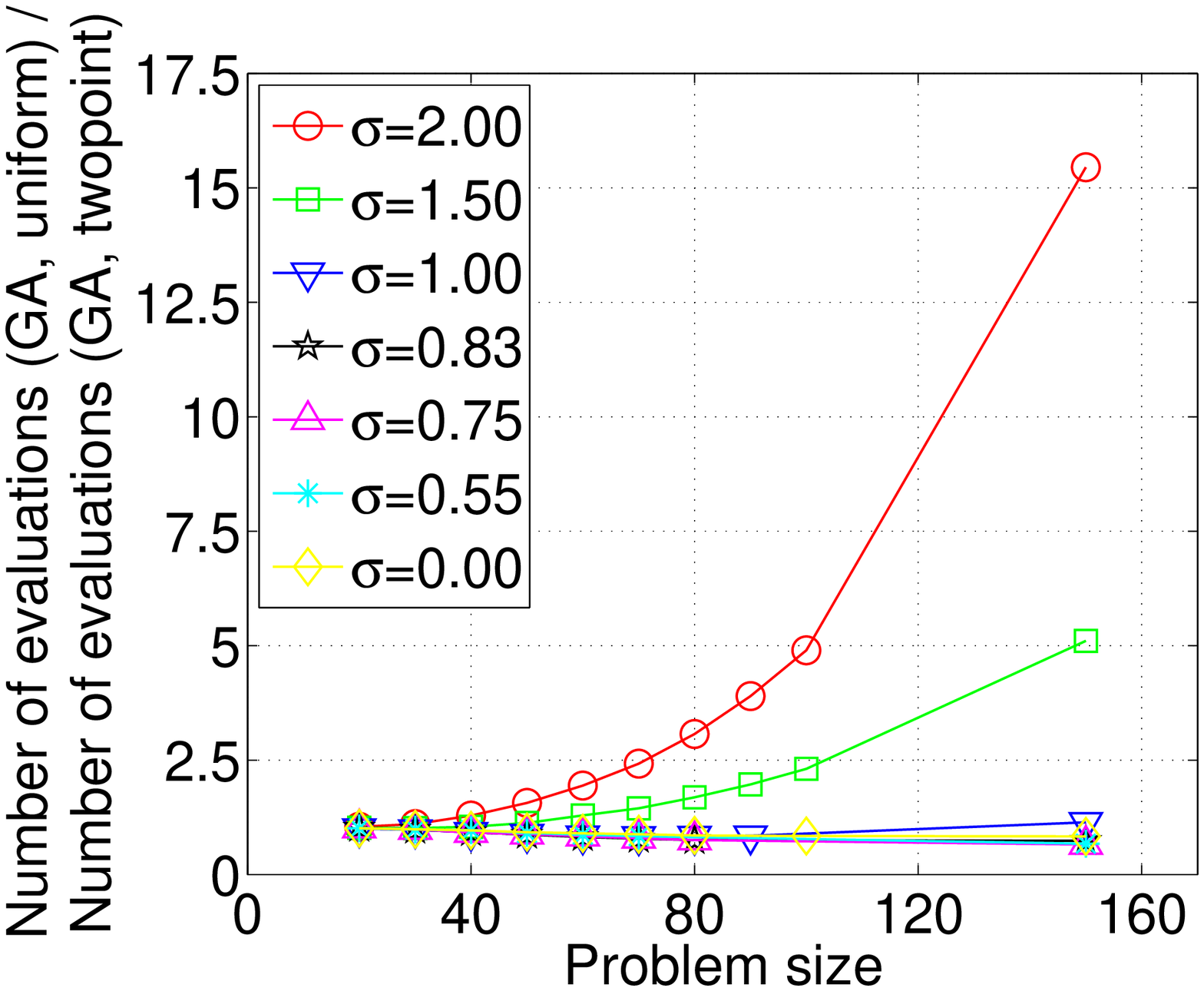,width=0.31 \textwidth}}
\hfill ~
\caption{Ratio for the number of evaluations for pairs of compared algorithms.}
\label{fig-ratio-evaluations}
\end{figure*}

\begin{figure*}[t]
\hfill ~
\subfigure[GA (twopoint) and hBOA]{\epsfig{file=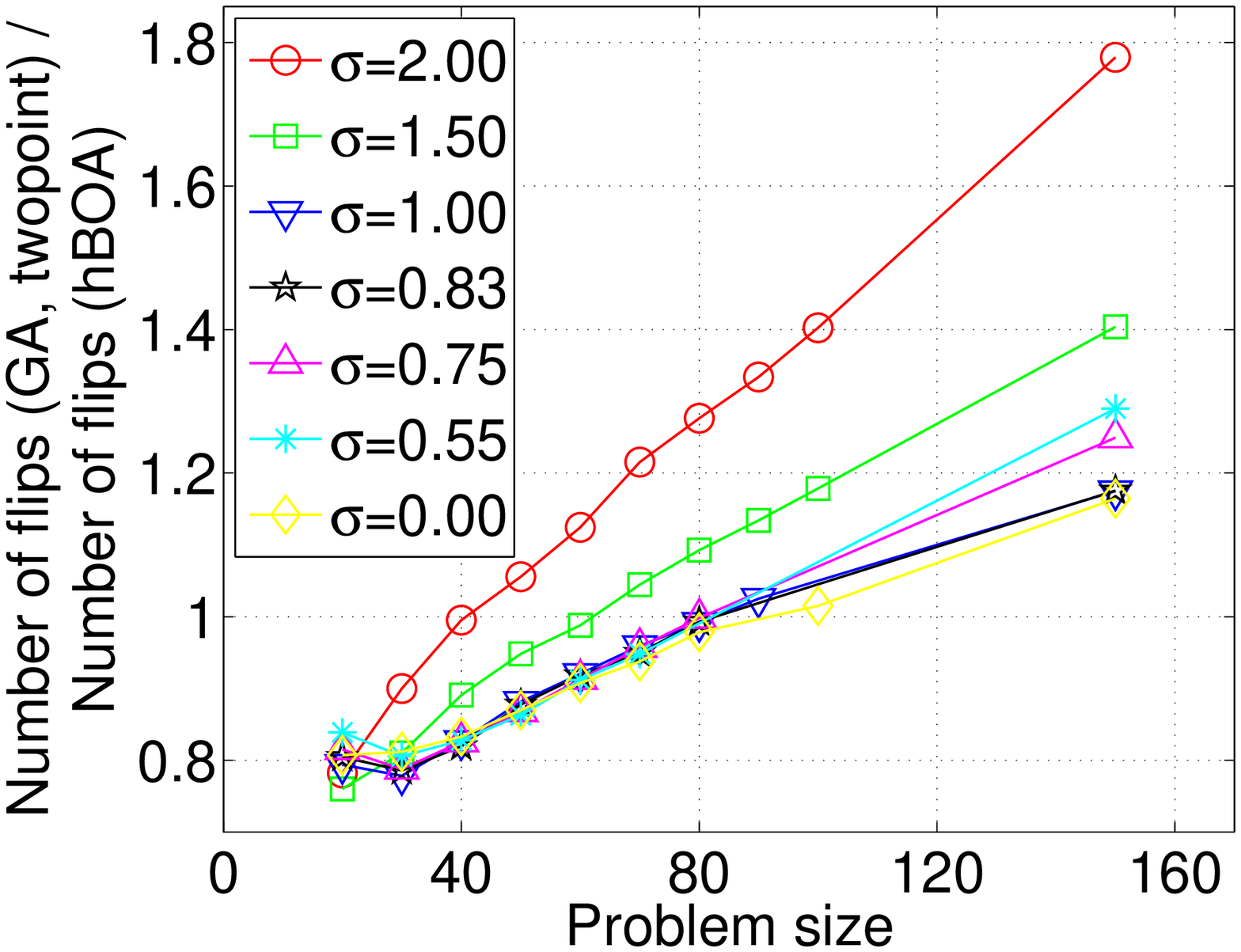,width=0.31 \textwidth}}
\hfill ~
\subfigure[GA (uniform) and hBOA]{\epsfig{file=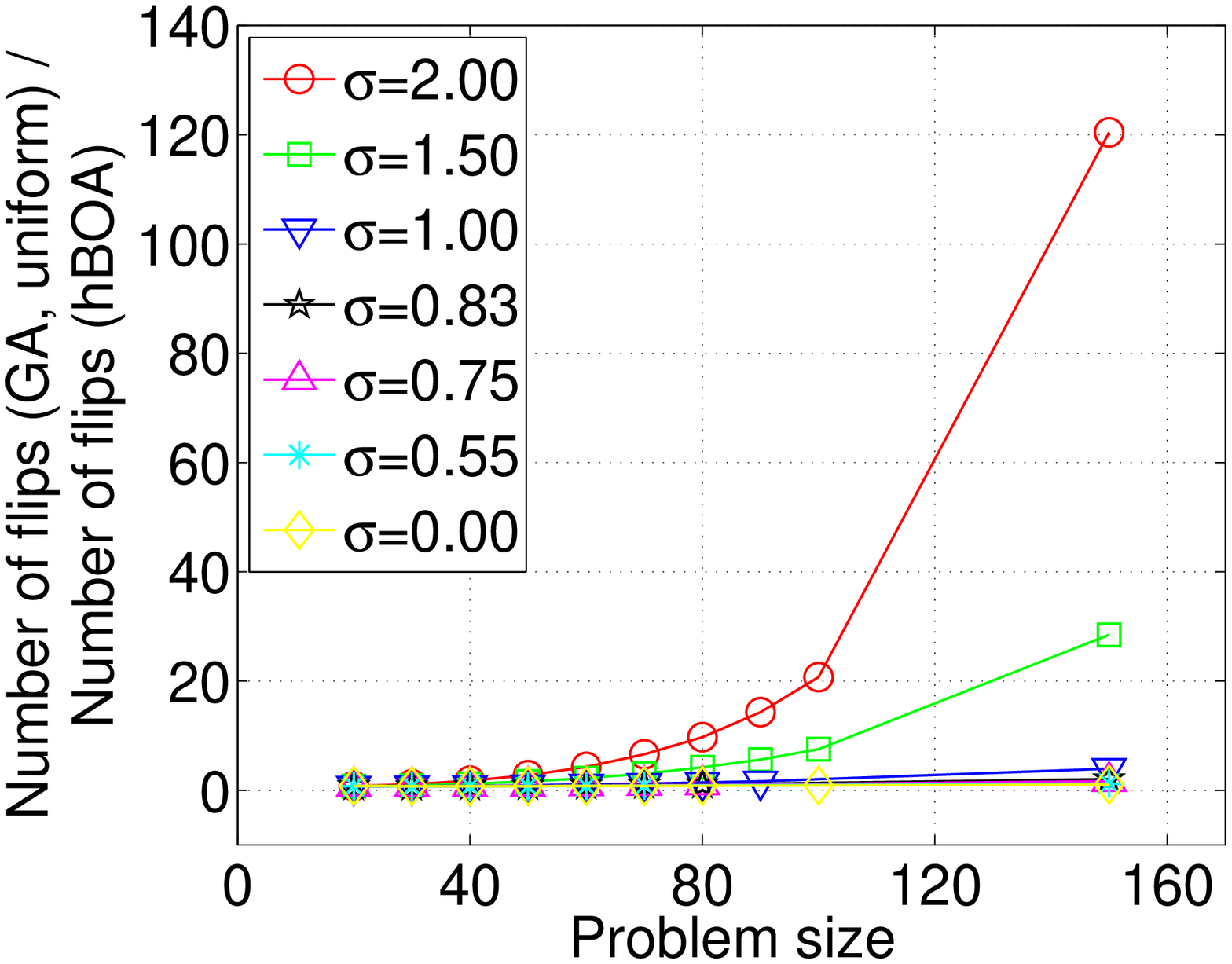,width=0.31 \textwidth}}
\hfill ~
\subfigure[GA (uniform) and GA (twopoint)]{\epsfig{file=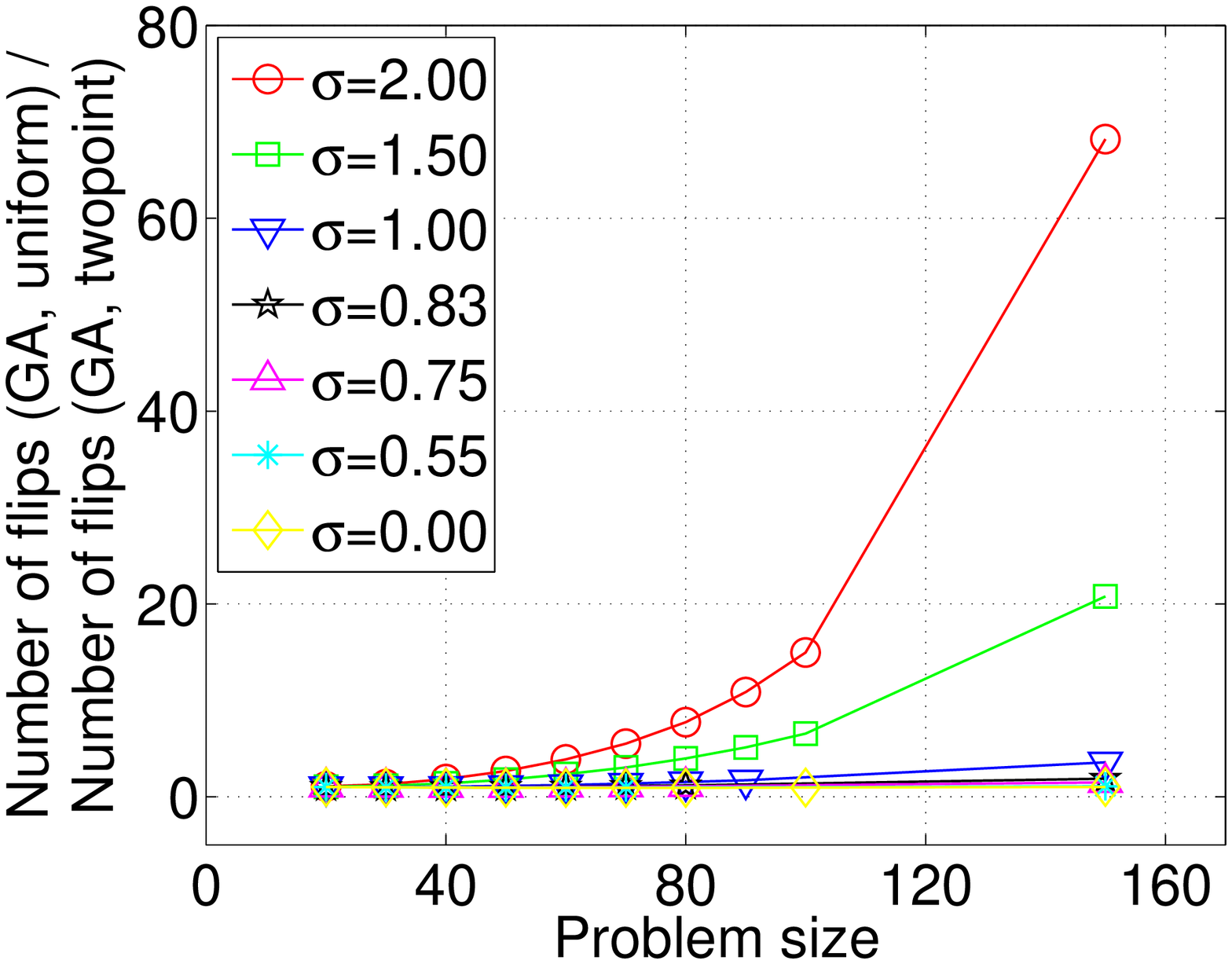,width=0.31 \textwidth}}
\hfill ~
\caption{Ratio for the number of flips for pairs of compared algorithms. }
\label{fig-ratio-flips}
\end{figure*}

\begin{figure*}[t]
\hfill ~
\subfigure[GA (twopoint) and hBOA]{\epsfig{file=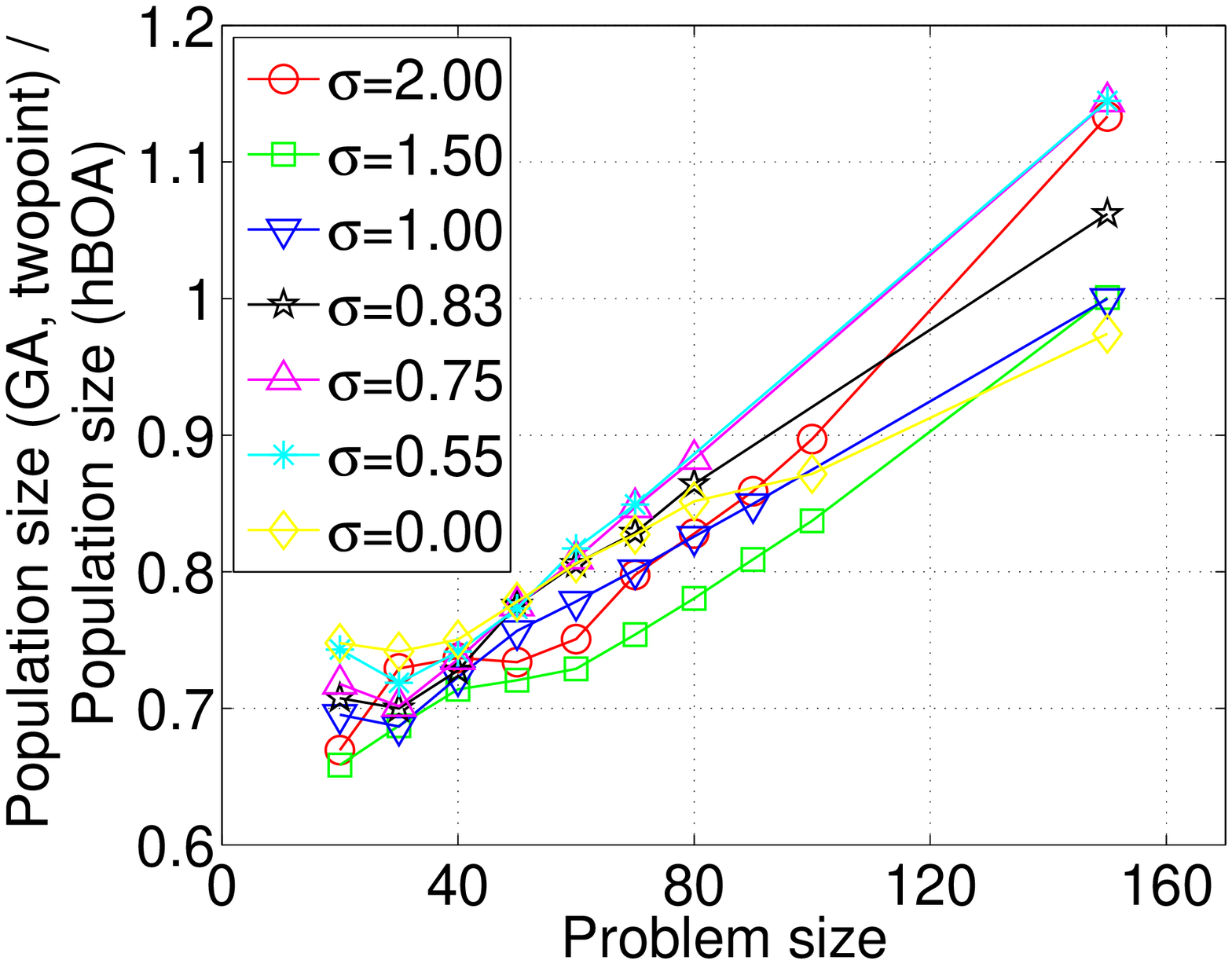,width=0.31 \textwidth}}
\hfill ~
\subfigure[GA (uniform) and hBOA]{\epsfig{file=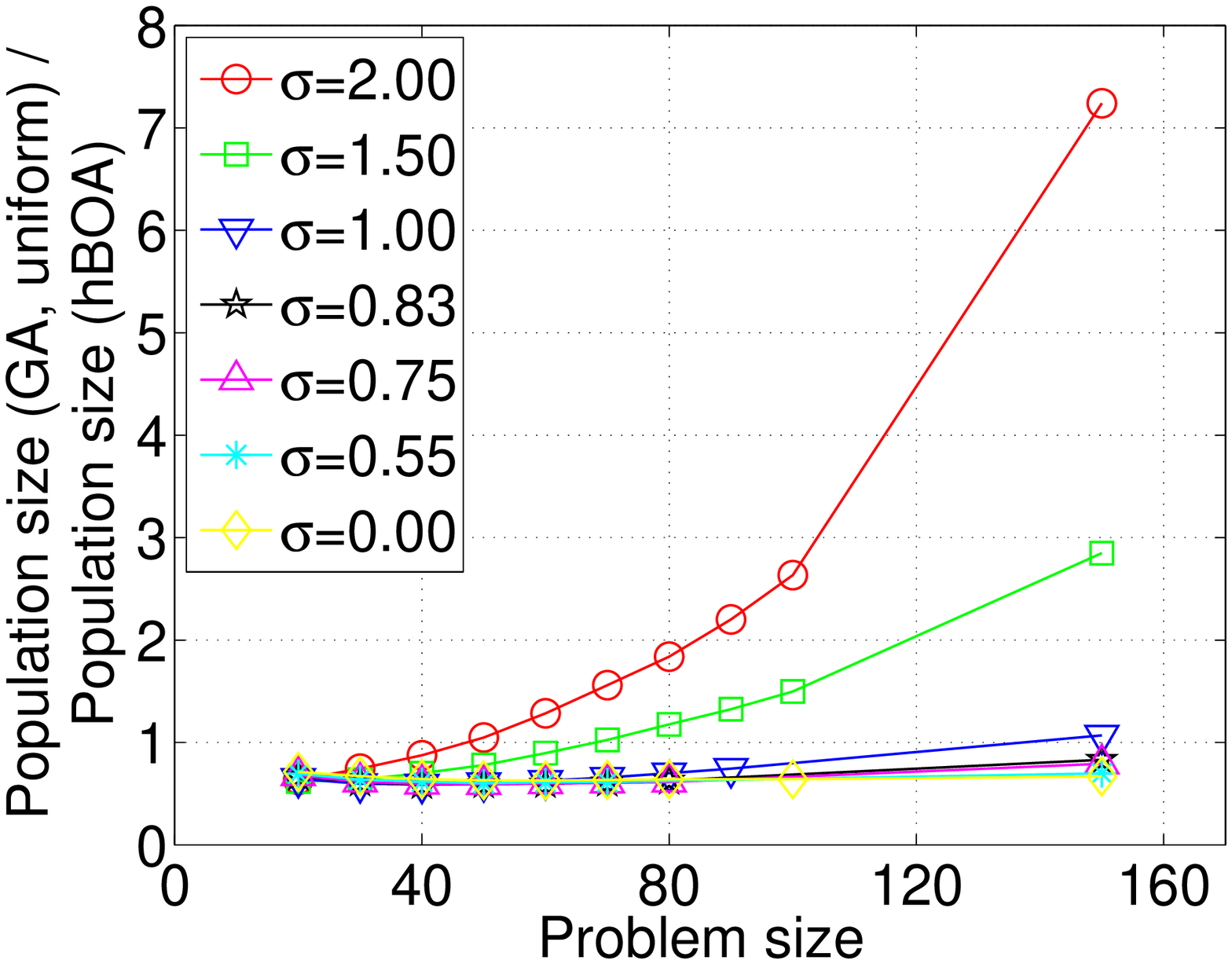,width=0.31 \textwidth}}
\hfill ~
\subfigure[GA (uniform) and GA (twopoint)]{\epsfig{file=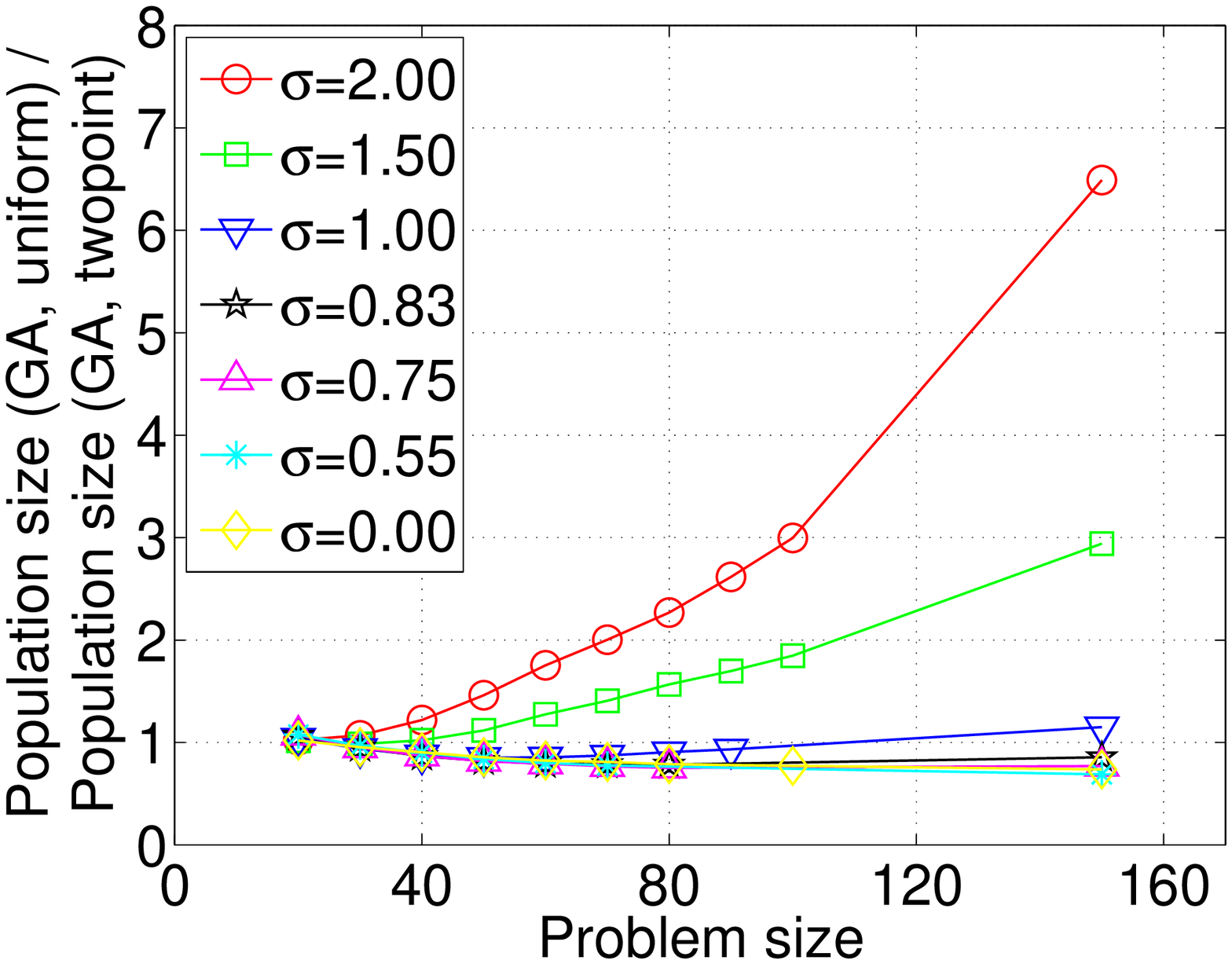,width=0.31 \textwidth}}
\hfill ~
\caption{Ratio for the required population sizes for pairs of compared algorithms. }
\label{fig-ratio-popsizes}
\end{figure*}

One of the interesting questions from the perspective of problem difficulty is whether the problem instances that are difficult for one algorithm are also difficult for other algorithms. One of the approaches to visualizing this is to look at the correlation between the running times between pairs of algorithms over problem instances for specific values of $n$ and $\sigma$. Figure~\ref{fig-correlation-flips} compares the running times measured by the number of flips for the largest problem of $n=150$ spins and $\sigma\in\{0.55,2.00\}$. The results indicate, both numerically and visually, the the running times for $\sigma=0.55$ are more strongly correlated than the running times for $\sigma=2.00$. A somewhat more detailed view on the correlation between the numbers of flips until optimum for the different algorithms is shown in figure~\ref{fig-correlation-with-sigma}, which confirms that the results are more strongly correlated for smaller values of $\sigma$; additionally, these results indicate that the correlations are almost consistently stronger for smaller problem sizes.

\begin{figure*}[t]
\centering
\subfigure[$\sigma=0.55$]{\epsfig{file=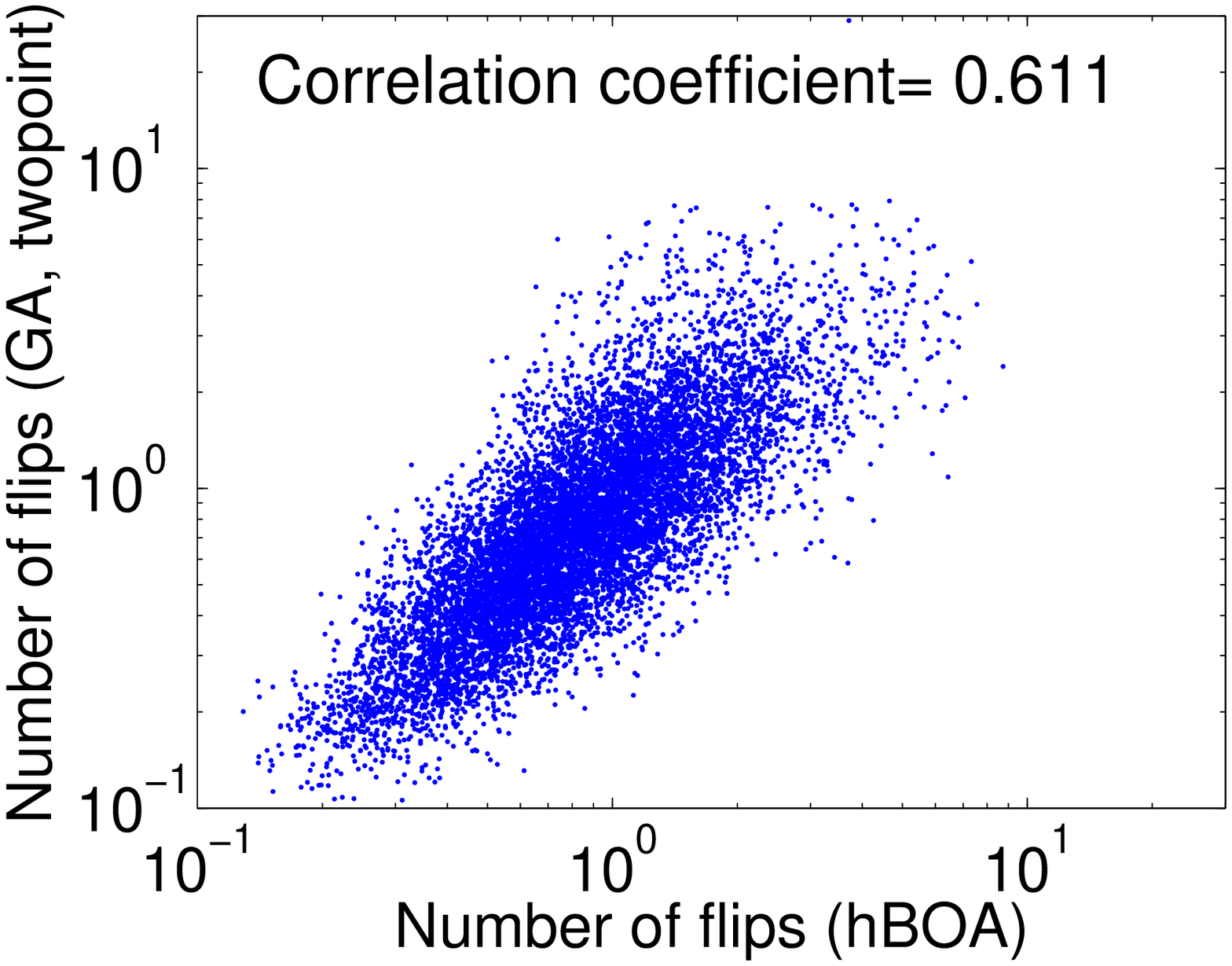,width=0.31 \textwidth}~~~
\epsfig{file=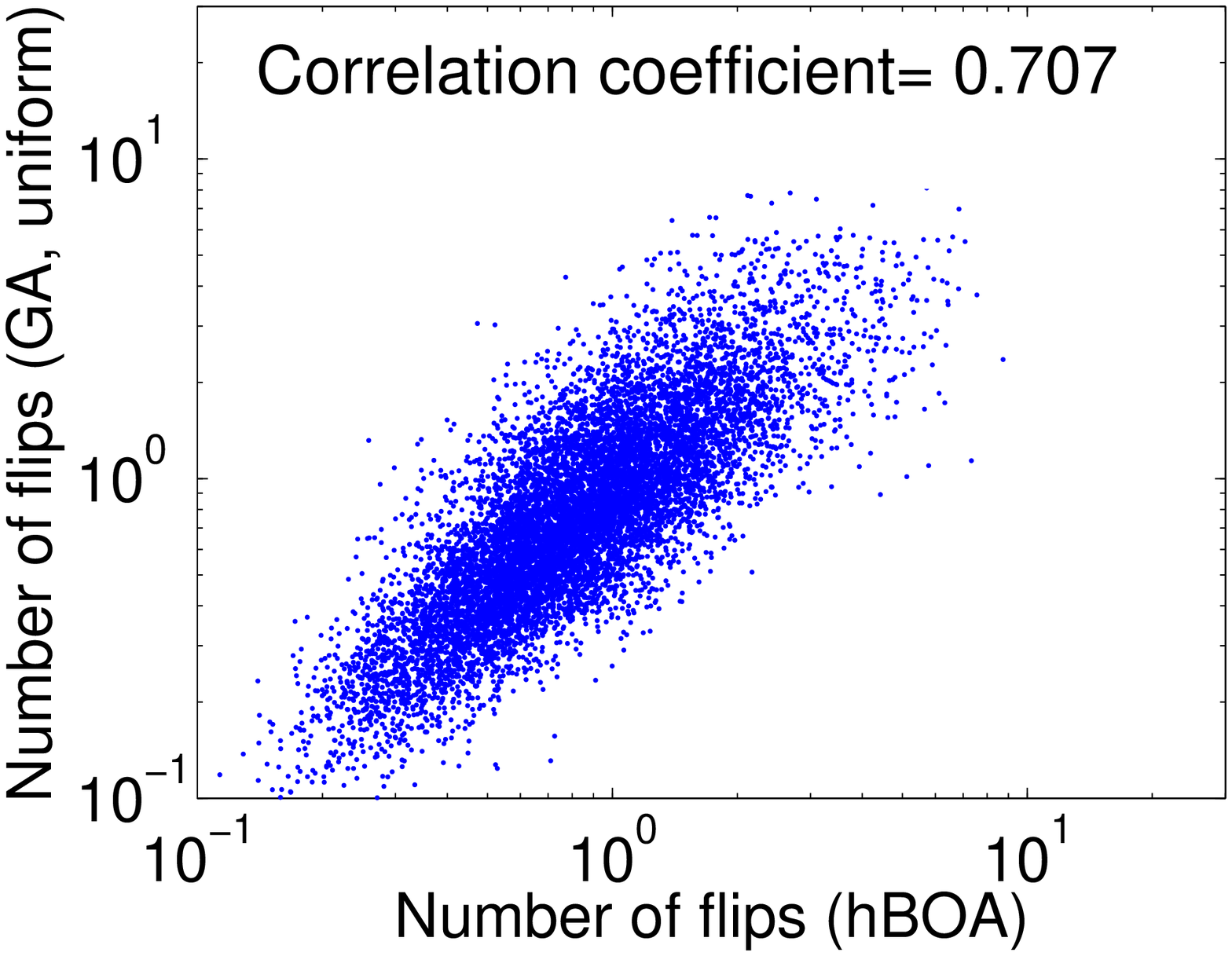,width=0.31 \textwidth}~~~
\epsfig{file=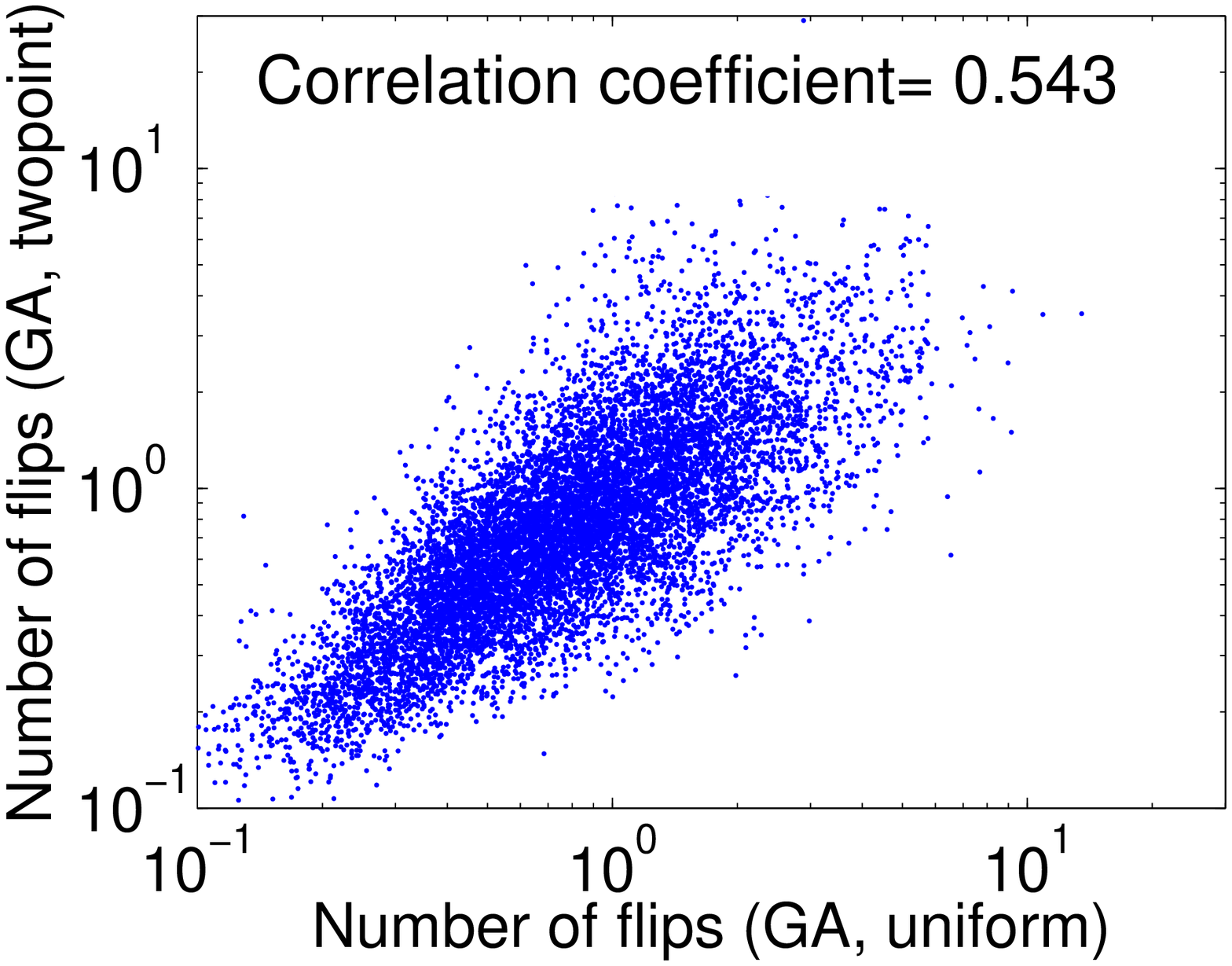,width=0.31 \textwidth}}
\\
\subfigure[$\sigma=2.00$]{\epsfig{file=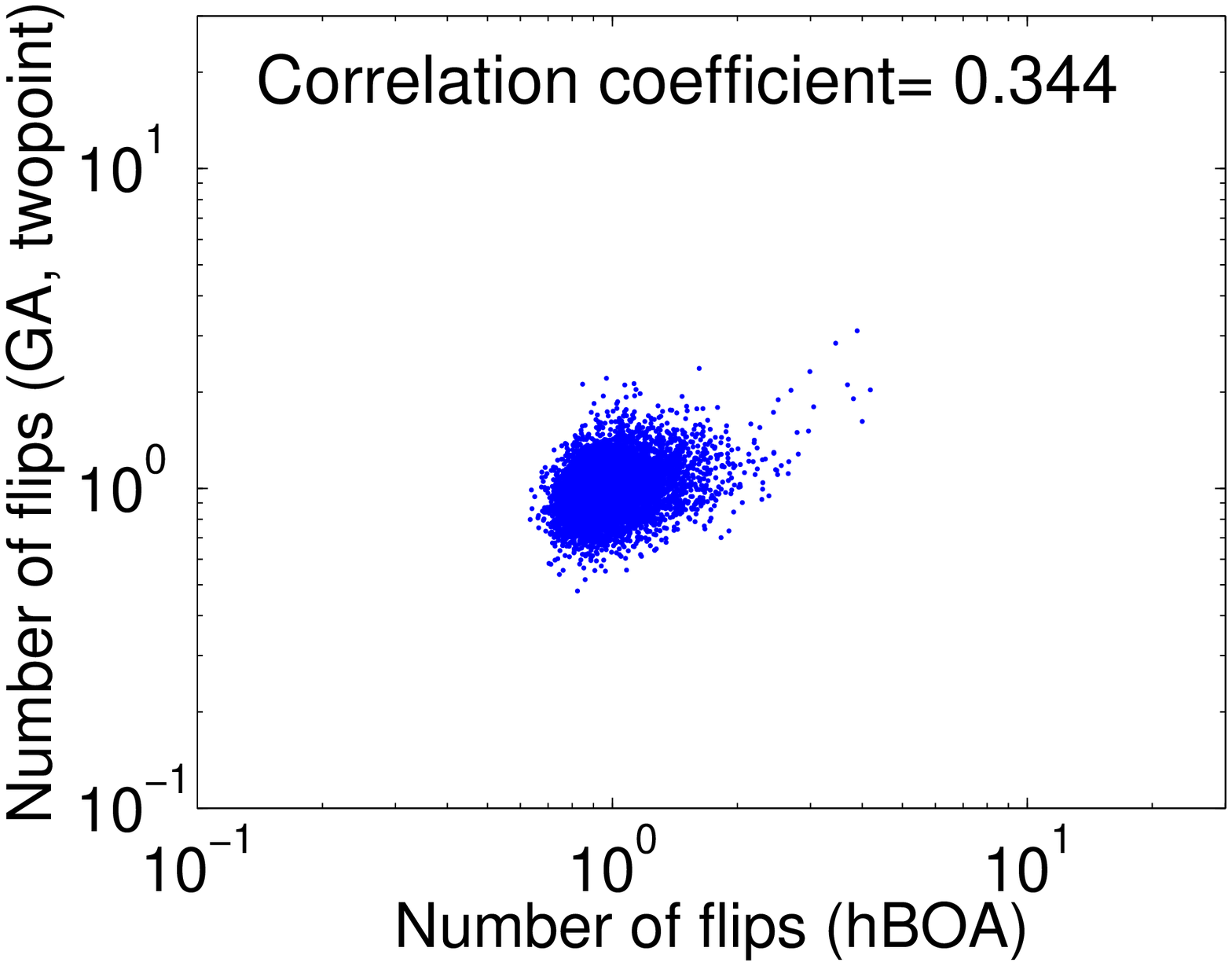,width=0.31 \textwidth}~~~
\epsfig{file=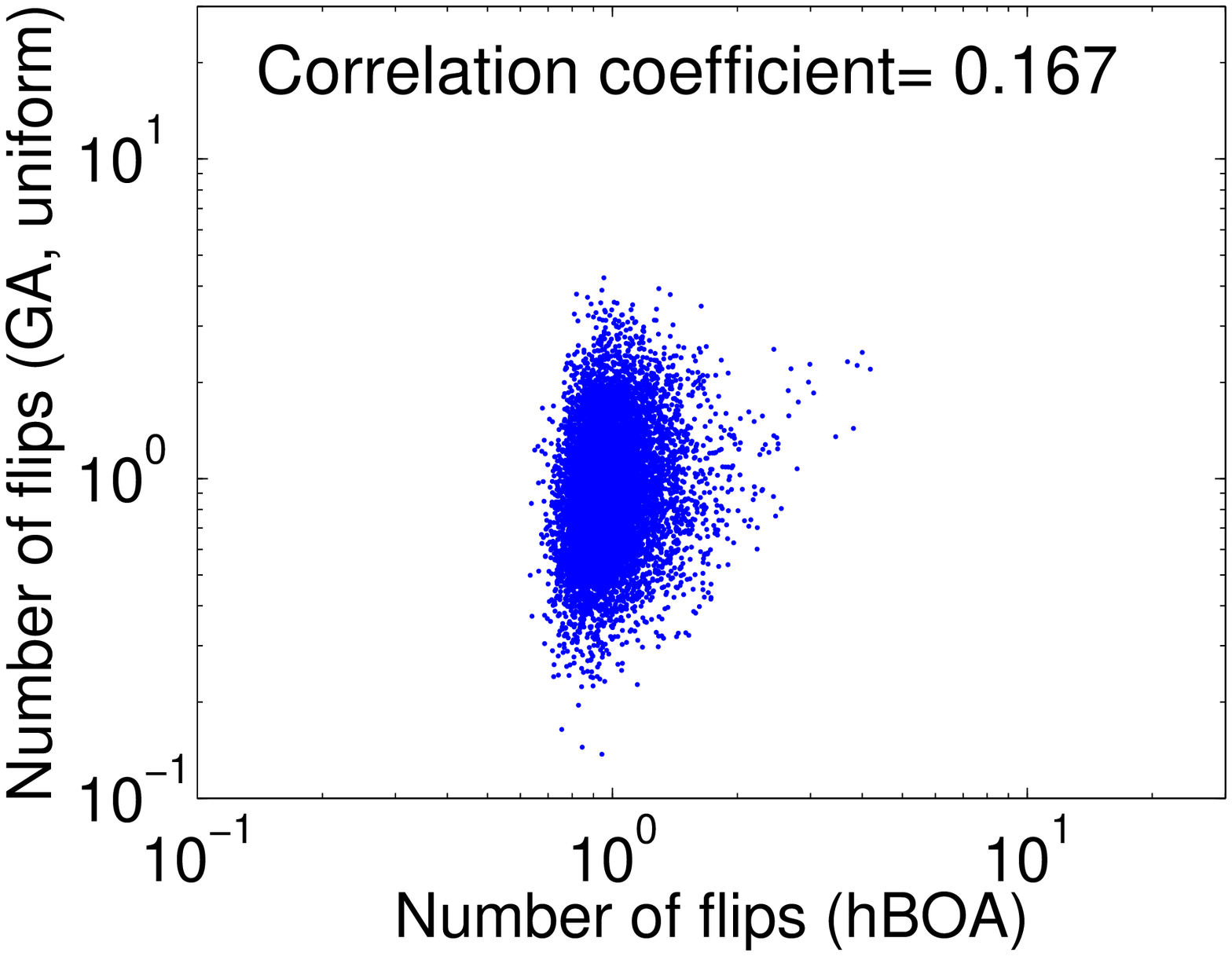,width=0.31 \textwidth}~~~
\epsfig{file=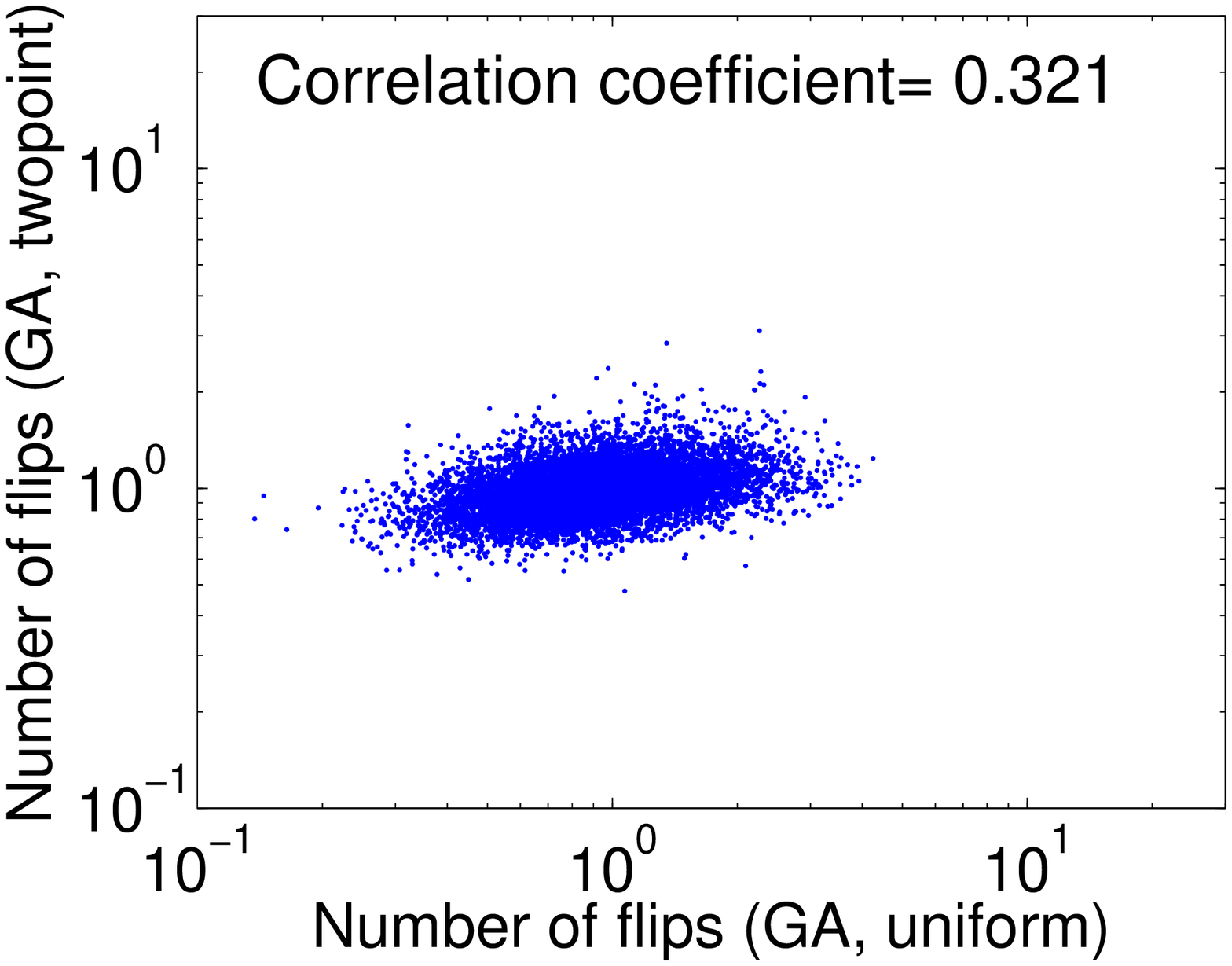,width=0.31 \textwidth}}
\caption{The correlation between the number of flips for different algorithms for $n=150$ and $\sigma\in \{0.55,2.00\}$.}
\label{fig-correlation-flips}
\end{figure*}

\begin{figure*}
\hfill ~
\subfigure[hBOA vs. GA (twopoint)]{\epsfig{file=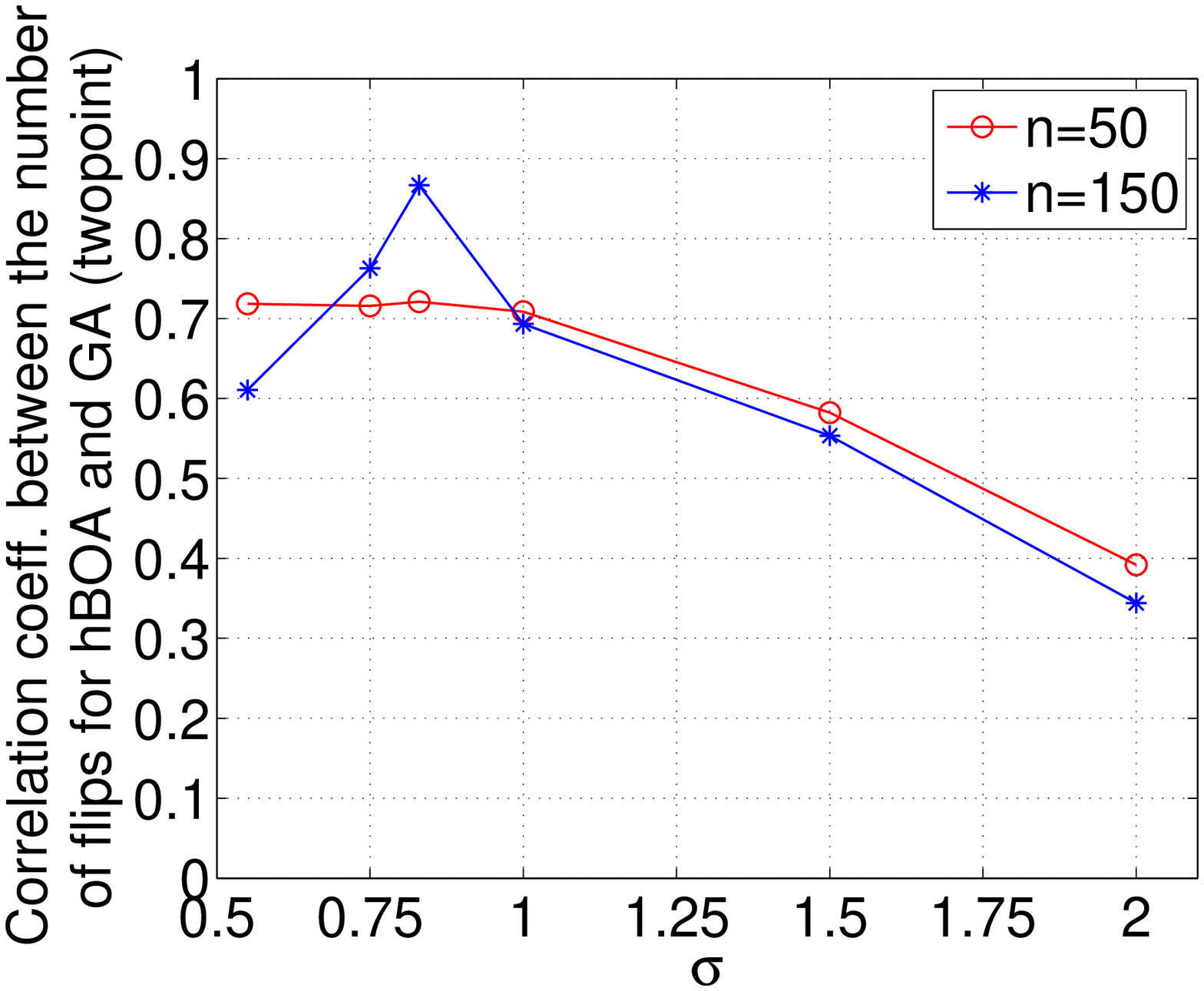,width=0.31\textwidth}}
\hfill ~
\subfigure[hBOA vs. GA (uniform)]{\epsfig{file=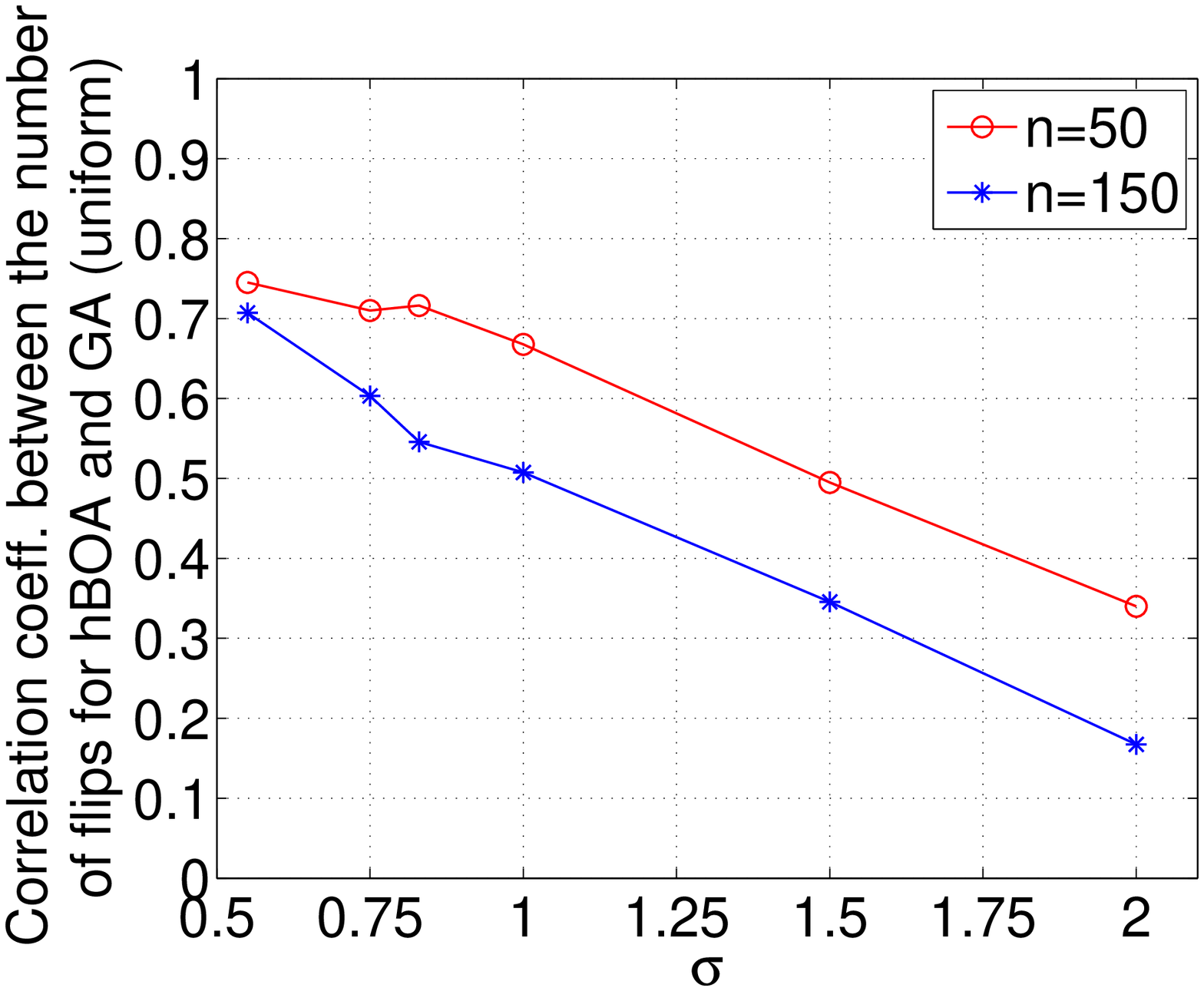,width=0.31\textwidth}}
\hfill ~
\subfigure[GA (uniform) vs. GA (twopoint)]{\epsfig{file=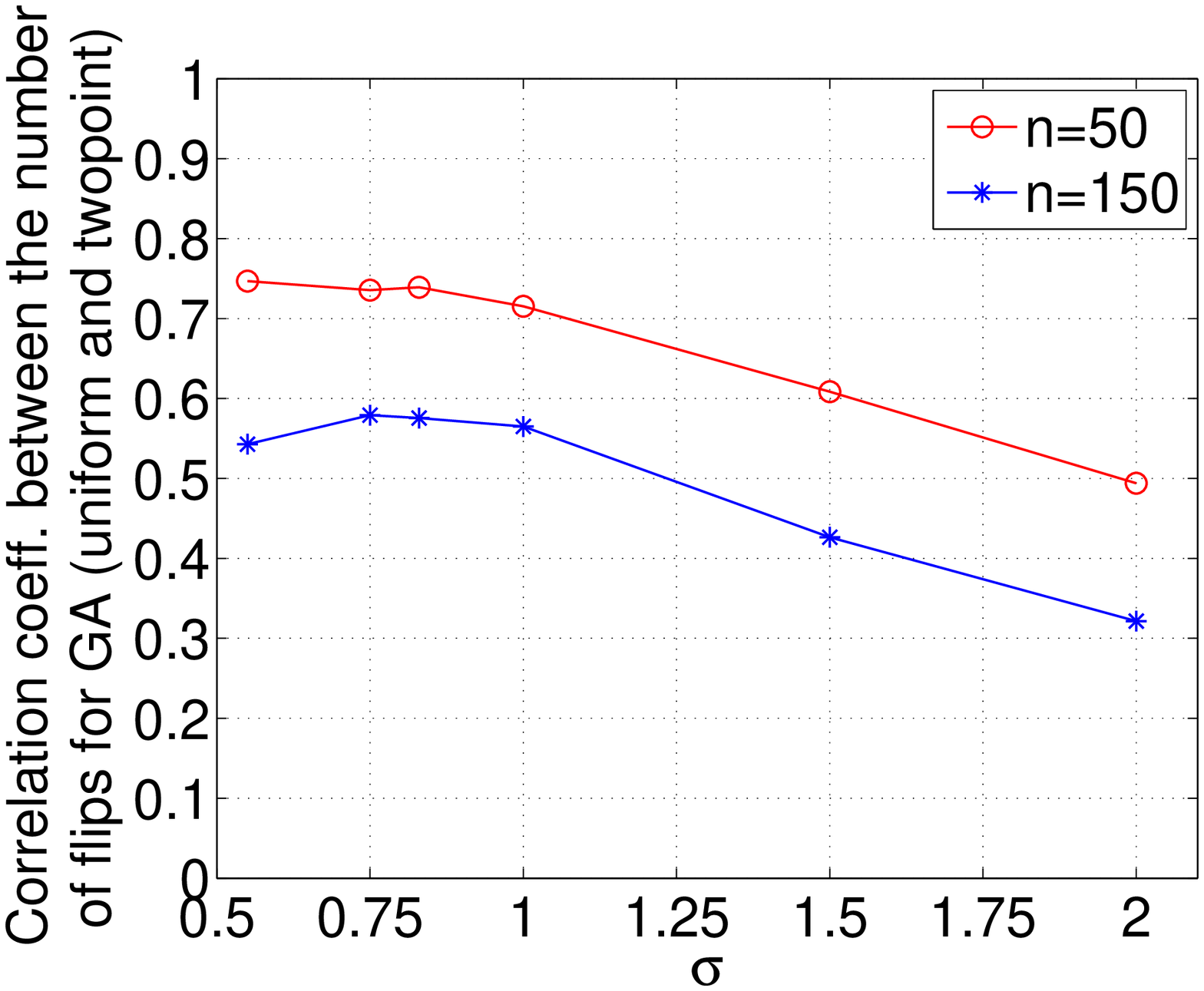,width=0.31\textwidth}}
\hfill ~
\caption{The correlation between the number of flips for different algorithms for $n=50$ and $n=150$.}
\label{fig-correlation-with-sigma}
\end{figure*}

\section{Future Work}
\label{section-future-work}
The numerous statistics provided by the experiments presented in this paper can be analyzed in more detail to provide a deeper insight into the strengths and weaknesses of the compared evolutionary algorithms, and to provide inputs for further improvement of these algorithms on the one-dimensional spin glass with power-law interactions and on other difficult classes of additively decomposable problems with complex fitness landscape.

One of the important outputs of this paper was the large number of problem instances of the one-dimensional spin glass with power-law interactions. These instances can be used for additional empirical studies in evolutionary computation and beyond. Most importantly, one should consider techniques that were shown to perform well on various classes of spin glass models and compare the performance of these techniques to that of the evolutionary algorithms studied in this paper. Three algorithms are of special interest: extremal optimization~\cite{Boettcher:05}, hysteretic optimization~\cite{Pal:96,Pal:06}, and GA with triadic crossover~\cite{Pal:96}. Hybrids of evolutionary algorithms, hysteretic optimization, and extremal optimization hold a great promise for solving large instances of both the general SK spin glass as well as the one-dimensional SK spin glass with power-law interactions. The key is to identify the strengths and weaknesses of the different algorithms and to combine their strengths effectively. 

Another interesting topic for future research is related to problem difficulty of the one-dimensional spin glass with power-law interactions and the general SK spin glass.  Is there a way to quantify features that make one problem instance more difficult than the other one besides the direct influence of $n$ and $\sigma$? Can scalability theory for evolutionary algorithms~\cite{Goldberg:91j,Goldberg:92c,Harik:97a**,Thierens:98} and estimation of distribution algorithms~\cite{Muhlenbein:93c,Muhlenbein:99,Pelikan:02a,Pelikan:book,Zhang:04b,Zhang:04c,Yu:07} be used to gain better understanding of problem difficulty of the considered spin glass model? Can problem difficulty be explained using existing techniques for landscape analysis in evolutionary computation and optimization theory~\cite{Naudts:97,Kallel_et_al:2000,Rothlauf:2002,Barnes:03} or do new tools have to be designed for this purpose? 

\section{Summary and Conclusions}
\label{section-conclusions}
The problem of finding ground states of the one-dimensional spin glass model with power-law interactions provides an interesting class of optimization benchmarks with tunable range of interactions. This paper presented an in-depth empirical analysis of the genetic algorithm (GA) with twopoint and uniform crossover and the hierarchical Bayesian optimization algorithm (hBOA) on this class of problems. A large number of instances of different size and range of interactions were generated. The branch and bound algorithm was then used to find guaranteed ground states of most instances; for largest instances, which were unsolvable in practical time by the branch and bound, the population-doubling approach based on hBOA was used to find reliable (but not guaranteed) optima. Overall, 610,000 unique problem instances were used. All algorithms were then applied to the generated instances, and the results were analyzed and discussed. 

The best performance was achieved by hBOA, the second best performance was achieved by GA with twopoint crossover and the worst performance was achieved by GA with uniform crossover. The differences between all algorithms became more substantial as the range of interactions was reduced and as the problem size was increased.

The results clearly indicated that when the recombination operator is able either to learn linkage automatically (hBOA) or to preserve linkage between consequent bits (GA with twopoint crossover), instances with short-range interactions are clearly easier than those with long-range interactions. Nonetheless, for recombination that processes each bit independently regardless of its position in the solution strings (GA with uniform crossover), problem instances with long-range interactions are in fact easier than those with short-range interactions. 

For instances with short-range interactions, both hBOA and GA with twopoint crossover were empirically shown to require low-order polynomial time whether we measure time complexity through the number of evaluations or the number of flips of the local searcher. Nonetheless, for long-range interactions, scalability of both algorithms suffered and the time required to solve these problems appeared to grow faster than polynomially. For GA with uniform crossover, the growth of the time complexity regardless of the range of interactions appeared to be faster than polynomial for the entire range of problem instances under consideration.

This paper presented the first empirical analysis of evolutionary algorithm on the problem of finding ground states of the one-dimensional spin glass with power-law interactions. We believe that this class of problems provides a rich source of problem instances for future research in evolutionary computation, both in the area of empirical testing as well as in the design of theoretical models of evolutionary algorithms on additively decomposable problems with a complex fitness landscape. 

\section*{Acknowledgments}
This project was sponsored by the National Science Foundation under CAREER grant ECS-0547013, by the Air Force Office of Scientific Research, USAF, under grant FA9550-06-1-0096, and by the University of Missouri in St. Louis through the High Performance Computing Collaboratory sponsored by Information Technology Services, and the Research Award and Research Board programs. The U.S.  Government is authorized to reproduce and distribute reprints for government purposes notwithstanding any copyright notation thereon. Any opinions, findings, and conclusions or recommendations expressed in this material are those of the authors and do not necessarily reflect the views of the National Science Foundation, the Air Force Office of Scientific Research, or the U.S. Government. Some experiments were done using the hBOA software developed by M. Pelikan and D. E. Goldberg at the Univ. of Illinois at Urbana-Champaign. H.G.K.~would like to thank the Swiss National Science Foundation for financial support under grant No.~PP002-114713. Most simulations were performed on the ETH Z\"{u}rich cluster and the Beowulf cluster of the University of Missouri in St. Louis.


\bibliographystyle{abbrv}


\begin{thebibliography}{10}

\bibitem{Baluja:94}
S.~Baluja.
\newblock Population-based incremental learning: {A} method for integrating
  genetic search based function optimization and competitive learning.
\newblock Tech. Rep. No. CMU-CS-94-163, Carnegie Mellon University, Pittsburgh,
  PA, 1994.

\bibitem{Barahona:82}
F.~Barahona.
\newblock On the computational complexity of {I}sing spin glass models.
\newblock {\em Journal of Physics A: Mathematical, Nuclear and General},
  15(10):3241--3253, 1982.

\bibitem{Barnes:03}
J.~W. Barnes, B.~Dimova, S.~P. Dokov, and A.~Solomon.
\newblock The theory of elementary landscapes.
\newblock {\em Appl. Math. Lett.}, 16(3):337--343, 2003.

\bibitem{Boettcher:05}
S.~Boettcher.
\newblock Extremal optimization for {S}herrington-{K}irkpatrick spin glasses.
\newblock {\em Eur. Phys. J. B}, 46:501--505, 2005.

\bibitem{Boettcher:08}
S.~Boettcher, H.~G. Katzgraber, and D.~Sherrington.
\newblock Local field distributions in spin glasses.
\newblock {\em Journal of Physics A: Mathematical and Theoretical},
  41(32):324007, 2008.

\bibitem{Bosman:00*}
P.~A.~N. Bosman and D.~Thierens.
\newblock Continuous iterated density estimation evolutionary algorithms within
  the {IDEA} framework.
\newblock {\em Workshop Proc. of the {G}enetic and {E}volutionary {C}omputation
  {C}onf. ({GECCO}-2000)}, pages 197--200, 2000.

\bibitem{Chickering:97}
D.~M. Chickering, D.~Heckerman, and C.~Meek.
\newblock A {B}ayesian approach to learning {B}ayesian networks with local
  structure.
\newblock Technical Report MSR-TR-97-07, Microsoft Research, Redmond, WA, 1997.

\bibitem{Wegener:2004}
S.~Fischer and I.~Wegener.
\newblock The ising model on the ring: Mutation versus recombination.
\newblock {\em Proc. of the {G}enetic and {E}volutionary {C}omputation {C}onf.
  ({GECCO}-2004)}, pages 1113--1124, 2004.

\bibitem{fisher:88}
D.~S. Fisher and D.~A. Huse.
\newblock Equilibrium behavior of the spin-glass ordered phase.
\newblock {\em Phys. Rev. B}, 38:386, 1988.

\bibitem{Friedman:99}
N.~Friedman and M.~Goldszmidt.
\newblock Learning {B}ayesian networks with local structure.
\newblock In M.~I. Jordan, editor, {\em Graphical models}, pages 421--459. MIT
  Press, Cambridge, MA, 1999.

\bibitem{Goldberg:89d}
D.~E. Goldberg.
\newblock {\em Genetic algorithms in search, optimization, and machine
  learning}.
\newblock Addison-Wesley, Reading, MA, 1989.

\bibitem{Goldberg:92c}
D.~E. Goldberg, K.~Deb, and J.~H. Clark.
\newblock Genetic algorithms, noise, and the sizing of populations.
\newblock {\em Complex {S}ystems}, 6:333--362, 1992.

\bibitem{Goldberg:91j}
D.~E. Goldberg and M.~Rudnick.
\newblock Genetic algorithms and the variance of fitness.
\newblock {\em Complex Systems}, 5(3):265--278, 1991.

\bibitem{Harik:95a}
G.~R. Harik.
\newblock Finding multimodal solutions using restricted tournament selection.
\newblock {\em Proc. of the {I}nternational {C}onf. on {G}enetic {A}lgorithms
  ({ICGA}-95)}, pages 24--31, 1995.

\bibitem{Harik:97a**}
G.~R. Harik, E.~Cant{\'{u}}-Paz, D.~E. Goldberg, and B.~L. Miller.
\newblock The gambler's ruin problem, genetic algorithms, and the sizing of
  populations.
\newblock {\em Proc. of the International Conf. on Evolutionary Computation
  ({ICEC}-97)}, pages 7--12, 1997.

\bibitem{Hartmann:01}
A.~K. Hartmann.
\newblock Ground-state clusters of two, three and four-dimensional +/-{J}
  {I}sing spin glasses.
\newblock {\em Phys. Rev. E}, 63:016106, 2001.

\bibitem{Kobe:84}
A.~Hartwig, F.~Daske, and S.~Kobe.
\newblock A recursive branch-and-bound algorithm for the exact ground state of
  {I}sing spin-glass models.
\newblock {\em Computer Physics Communications}, 32:133--138, 1984.

\bibitem{Holland:75a}
J.~H. Holland.
\newblock {\em Adaptation in natural and artificial systems}.
\newblock University of Michigan Press, Ann Arbor, MI, 1975.

\bibitem{Kallel_et_al:2000}
L.~Kallel, B.~Naudts, and R.~Reeves.
\newblock Properties of fitness functions and search landscapes.
\newblock In L.~Kallel, B.~Naudts, and A.~Rogers, editors, {\em Theoretical
  Aspects of Evolutionary Computing}, pages 177--208. Springer Verlag, 2000.

\bibitem{Katzgraber:08}
H.~G. Katzgraber.
\newblock Spin glasses and algorithm benchmarks: A one-dimensional view.
\newblock {\em Journal of Physics: Conf. Series}, 95:012004, 2008.

\bibitem{katzgraber:04c}
H.~G. Katzgraber, M.~{K{\"o}rner}, F.~{Liers}, M.~{J{\"u}nger}, and A.~K.
  {Hartmann}.
\newblock {{Universality-class dependence of energy distributions in spin
  glasses}}.
\newblock {\em Phys. Rev. B}, 72:094421, 2005.

\bibitem{katzgraber:03}
H.~G. Katzgraber and A.~P. Young.
\newblock Monte {C}arlo studies of the one-dimensional {I}sing spin glass with
  power-law interactions.
\newblock {\em Phys. Rev. B}, 67:134410, 2003.

\bibitem{Sherrington:78}
S.~Kirkpatrick and D.~Sherrington.
\newblock Infinite-ranged models of spin-glasses.
\newblock {\em Phys. Rev. B}, 17(11):4384--4403, Jun 1978.

\bibitem{Kobe:03}
S.~Kobe.
\newblock Ground-state energy and frustration of the
  {S}herrington-{K}irkpatrick model and related models.
\newblock ArXiv Condensed Matter e-print cond-mat/03116570, University of
  Dresden, 2003.

\bibitem{kotliar:83}
G.~{Kotliar}, P.~W. {Anderson}, and D.~L. {Stein}.
\newblock One-dimensional spin-glass model with long-range random interactions.
\newblock {\em Phys. Rev. B}, 27:R602, 1983.

\bibitem{Larranaga:02}
P.~Larra{\~{n}}aga and J.~A. Lozano, editors.
\newblock {\em Estimation of Distribution Algorithms: A New Tool for
  Evolutionary Computation}.
\newblock Kluwer, Boston, MA, 2002.

\bibitem{Larranaga:06}
J.~A. Lozano, P.~Larra{\~{n}}aga, I.~Inza, and E.~Bengoetxea, editors.
\newblock {\em Towards a New Evolutionary Computation: Advances on Estimation
  of Distribution Algorithms}.
\newblock Springer, 2006.

\bibitem{Muhlenbein:98a}
H.~M{\"{u}}hlenbein and T.~Mahnig.
\newblock Convergence theory and applications of the factorized distribution
  algorithm.
\newblock {\em Journal of Computing and Information Technology}, 7(1):19--32,
  1998.

\bibitem{Muhlenbein:99}
H.~M{\"{u}}hlenbein and T.~Mahnig.
\newblock {FDA} -- {A} scalable evolutionary algorithm for the optimization of
  additively decomposed functions.
\newblock {\em Evolutionary Computation}, 7(4):353--376, 1999.

\bibitem{Muhlenbein:96**}
H.~M{\"{u}}hlenbein and G.~Paa{\ss}.
\newblock From recombination of genes to the estimation of distributions {I}.
  {B}inary parameters.
\newblock {\em Parallel {P}roblem {S}olving from {N}ature}, pages 178--187,
  1996.

\bibitem{Muhlenbein:93c}
H.~M{\"{u}}hlenbein and D.~Schlierkamp-{V}oosen.
\newblock Predictive models for the breeder genetic algorithm: {I}.
  {C}ontinuous parameter optimization.
\newblock {\em Evolutionary {C}omputation}, 1(1):25--49, 1993.

\bibitem{Naudts:98*}
B.~Naudts and J.~Naudts.
\newblock The effect of spin-flip symmetry on the performance of the simple
  {GA}.
\newblock {\em Parallel {P}roblem {S}olving from {N}ature}, pages 67--76, 1998.

\bibitem{Naudts:97}
B.~Naudts, D.~Suys, and A.~Verschoren.
\newblock Epistasis as a basic concept in formal landscape analysis.
\newblock {\em Proc. of the {I}nternational {C}onf. on {G}enetic {A}lgorithms
  ({ICGA}-97)}, pages 65--72, 1997.

\bibitem{Pal:96}
K.~F. P{\'{a}}l.
\newblock The ground state energy of the {E}dwards-{A}nderson {I}sing spin
  glass with a hybrid genetic algorithm.
\newblock {\em Physica A}, 223(3-4):283--292, 1996.

\bibitem{Pal:06}
K.~F. P{\'{a}}l.
\newblock Hysteretic optimization for the {S}herrington {K}irkpatrick spin
  glass.
\newblock {\em Physica A}, 367:261--268, 2006.

\bibitem{Pelikan:book}
M.~Pelikan.
\newblock {\em Hierarchical {B}ayesian optimization algorithm: {T}oward a new
  generation of evolutionary algorithms}.
\newblock Springer, 2005.

\bibitem{Pelikan:01*}
M.~Pelikan and D.~E. Goldberg.
\newblock Escaping hierarchical traps with competent genetic algorithms.
\newblock {\em Proc. of the {G}enetic and {E}volutionary {C}omputation {C}onf.
  ({GECCO}-2001)}, pages 511--518, 2001.

\bibitem{Pelikan:03*}
M.~Pelikan and D.~E. Goldberg.
\newblock Hierarchical {BOA} solves {I}sing spin glasses and maxsat.
\newblock {\em Proc. of the {G}enetic and {E}volutionary {C}omputation {C}onf.
  ({GECCO}-2003)}, {II}:1275--1286, 2003.

\bibitem{Pelikan:03b}
M.~Pelikan and D.~E. Goldberg.
\newblock A hierarchy machine: {L}earning to optimize from nature and humans.
\newblock {\em Complexity}, 8(5):36--45, 2003.

\bibitem{Pelikan:02}
M.~Pelikan, D.~E. Goldberg, and F.~Lobo.
\newblock A survey of optimization by building and using probabilistic models.
\newblock {\em Computational Optimization and Applications}, 21(1):5--20, 2002.

\bibitem{Pelikan:06c}
M.~Pelikan and A.~K. Hartmann.
\newblock Hierarchical {BOA}, cluster exact approximation, and {I}sing spin
  glasses.
\newblock {\em Parallel {P}roblem {S}olving from {N}ature}, pages 122--131,
  2006.

\bibitem{Pelikan:08}
M.~Pelikan, H.~G. Katzgraber, and S.~Kobe.
\newblock Finding ground states of sherrington-kirkpatrick spin glasses with
  hierarchical {BOA} and genetic algorithms.
\newblock {\em Proc. of the {G}enetic and {E}volutionary {C}omputation {C}onf.
  ({GECCO}-2008)}, pages 447--454, 2008.

\bibitem{Pelikan:EDA-book}
M.~Pelikan, K.~Sastry, and E.~Cant{\'{u}}-Paz, editors.
\newblock {\em Scalable optimization via probabilistic modeling: {F}rom
  algorithms to applications}.
\newblock Springer-Verlag, 2006.

\bibitem{Pelikan:02a}
M.~Pelikan, K.~Sastry, and D.~E. Goldberg.
\newblock Scalability of the {B}ayesian optimization algorithm.
\newblock {\em International Journal of Approximate Reasoning}, 31(3):221--258,
  2002.

\bibitem{Rothlauf:2002}
F.~Rothlauf.
\newblock {\em Representations for genetic and evolutionary algorithms}.
\newblock Springer Verlag, Berlin, 2002.

\bibitem{Sastry:01c}
K.~Sastry.
\newblock Evaluation-relaxation schemes for genetic and evolutionary
  algorithms.
\newblock Master's thesis, University of Illinois at Urbana-Champaign,
  Department of General Engineering, Urbana, IL, 2001.

\bibitem{Thierens:98}
D.~Thierens, D.~E. Goldberg, and A.~G. Pereira.
\newblock Domino convergence, drift, and the temporal-salience structure of
  problems.
\newblock {\em Proc. of the International Conf. on Evolutionary Computation
  ({ICEC}-98)}, pages 535--540, 1998.

\bibitem{Hoyweghen:01a}
C.~{Van Hoyweghen}.
\newblock Detecting spin-flip symmetry in optimization problems.
\newblock In L.~Kallel et~al., editors, {\em Theoretical Aspects of
  Evolutionary Computing}, pages 423--437. Springer, Berlin, 2001.

\bibitem{Yu:07}
T.-L. Yu, K.~Sastry, D.~E. Goldberg, and M.~Pelikan.
\newblock Population sizing for entropy-based model building in discrete
  estimation of distribution algorithms.
\newblock {\em Proc. of the {G}enetic and {E}volutionary {C}omputation {C}onf.
  ({GECCO}-2007)}, pages 601--608, 2007.

\bibitem{Zhang:04b}
Q.~Zhang.
\newblock On stability of fixed points of limit models of univariate marginal
  distribution algorithm and factorized distribution algorithm.
\newblock {\em IEEE Transactions on Evolutionary Computation}, 8:80--93, 2004.

\bibitem{Zhang:04c}
Q.~Zhang and H.~Muhlenbein.
\newblock On the convergence of a class of estimation of distribution
  algorithms.
\newblock {\em IEEE Transactions on Evolutionary Computation}, 8:127--136,
  2004.

\end{thebibliography}

\end{sloppy}
\end{document}